\documentclass[12pt]{article}

\usepackage{graphics}
\usepackage{amsthm,amssymb,mathrsfs,setspace,amsmath}
\numberwithin{equation}{section}
\usepackage{graphicx}
\usepackage{booktabs}
\usepackage{tikz}
\usepackage{multirow}
\usepackage{setspace}
\usepackage{subcaption}
\usepackage{comment}
\usepackage[english]{babel}

\title{ A Doubly-Flexible Model Based on Generalized Gamma Frailty for Two-component Load-sharing Systems }
\author{
Shilpi Biswas \thanks{Indian
Institute of Technology Guwahati, Assam 781039, India; Email:
shilpi.biswas@iitg.ac.in}, 
Ayon Ganguly \thanks{Indian
Institute of Technology Guwahati, Assam 781039, India; Email:
aganguly@iitg.ac.in},
Debanjan Mitra \thanks{Indian Institute of Management Udaipur, Rajasthan 313001,
India; Email: debanjan.mitra@iimu.ac.in}
}
\date{}

\allowdisplaybreaks

\begin{document}

\maketitle

\begin{abstract}
   For two-component load-sharing systems, a doubly-flexible model is developed where the generalized Fruend bivariate (GFB) distribution is used for the baseline of the component lifetimes, and the generalized gamma (GG) family of distributions is used to incorporate a shared frailty that captures dependence between the component lifetimes. The proposed model structure results in a very general two-way class of models that enables a researcher to choose an appropriate model for a given two-component load-sharing data within the respective families of distributions. The GFB-GG model structure provides better fit to two-component load-sharing systems compared to existing models. Fitting methods for the proposed model, based on direct optimization and an expectation maximization (EM) type algorithm, are discussed. Through simulations, effectiveness of the fitting methods is demonstrated. Also, through simulations, it is shown that the proposed model serves the intended purpose of model choice for a given two-component load-sharing data. A simulation case, and analysis of a real dataset are presented to illustrate the strength of the proposed model. 
\end{abstract}

\noindent


\section{Introduction}\label{sec:intro}
Modeling of data from load-sharing parallel systems is of interest in reliability analysis since the contribution of Daniels~\cite{Daniels1945} in the textile industry. In a load-sharing parallel system, components are attached in a parallel fashion, where the constant total load on the system is shared by its components. As and when components within the system fail one by one, the load on the remaining operational components increases, thereby increasing their chance of failure. Load-sharing systems are commonly observed in manufacturing, textile industry, software reliability~\cite{Park2010}, etc. 

The literature on load-sharing systems consists of models and related inferential methods of various types. In particular, parametric models have been extensively used in this context wherein the component lifetimes are assumed to follow some statistical parametric distribution(s). Some early major contributions in this area were by Freund~\cite{Freund1961}, and Birnbaum and Saunders~\cite{Birnbaum1958}, among others, who developed bivariate distributions from physical considerations of component lifetimes within a load-sharing system. 

Among more recent works, Kim and Kvam~\cite{KimKvam2004} is an important landmark. This is the first work that introduced unknown load-share rules and estimated load-share parameters from available data, assuming the component lifetimes to be exponentially distributed. Park~\cite{Park2010, Park2013} advanced this modeling approach further, by using distributions like Weibull and lognormal for component lifetimes. Through Asha et al.'s~\cite{AKK2016} extended Freund's bivariate (EFB) distribution, more flexibility came into the parametric modeling of two-component load-sharing systems. According to this model, the distribution of the lifetime of the second component changes after the first component failure by means of a new failure rate parameter, although the assumed baseline distribution for component lifetimes remains of the same form. Franco et al.~\cite{FVK2020} further generalized this model, and proposed the generalized Freund bivariate (GFB) distribution for load-sharing systems, wherein the baseline distribution of the component lifetimes can also change after the first component failure. It may be mentioned here that when assumptions of specific statistical distributions for component lifetimes are difficult to make, a semiparametric modeling approach, as discussed by Kvam and Pena~\cite{KvamPena2005}, may be used. To this effect, Deshpande et al.~\cite{DDN2010} used a general semiparametric family of distributions for $k$-out-of-$n$ load-sharing systems.  

One of the key aspects of modeling load-sharing data is to appropriately capture the dependence structure among the components. Towards this, various methods have been proposed. For example, models based on sequential order statistics for $k$-out-of-$n$ systems where the joint distribution of the component lifetimes conditional on the recent failures have been considered; see, for example, Cramer and Kamps \cite{CK1996, Cramer2001}. Further developments along this line, among others, were by Balakrishnan et al.~\cite{BBK2011}, who considered a link function among the model parameters, and Pesch et al.~\cite{Pesch2024}, who extended the approach to heterogeneous components.

Two significant recent additions to the literature are by Franco et al.~\cite{FVK2020} and Asha et al.~\cite{ARR2018}, who have used bivariate statistical distributions for modeling load-sharing data. The GFB distribution of Franco et al.~\cite{FVK2020} models the change in the component lifetime distribution after the first failure, considering its effect on the proportional failure rate (PFR) parameter of the remaining components. Here, it may be noted that a class of distributions is called a PFR class if its reliability function is given by $(R_B(y;\theta_B))^\alpha,~ \alpha>0$, with $R_B(\cdot;\theta_B)$ as the baseline reliability function involving parameter $\theta_B$. Here, $\alpha$ is called the PFR parameter. Note that in the PFR class, the hazard or failure rate is proportional to the hazard rate of the baseline distribution function. Examples of distributions in the PFR class are exponential, Weibull, gamma etc. The GFB distribution of Franco et al.~\cite{FVK2020} is a further generalization of the EFB distribution of Asha et al.~\cite{AKK2016}. On the other hand, Asha et al.~\cite{ARR2018} use a shared frailty model to accommodate unobserved random effect that induces dependence among the components, when the component lifetimes are Weibull distributed. In particular, they use a positive stable frailty distribution in the setting of the EFB distribution for two-component load-sharing systems that they proposed earlier~\cite{AKK2016, ARR2018}. 

A shared frailty is a random factor incorporated into a model to account for unobserved random effects that may be present in the study units due to various factors, for example, some commonality among them~\cite{Hougaard2000}. In reliability and survival studies, the literature on shared frailty models is well developed~\cite{Hanagal2019}. For load-sharing systems, a shared frailty can conveniently incorporate dependence between the component lifetimes~\cite{ARR2018}. The shared frailty model can also capture the unit-to-unit variation of the study units. However, since a shared frailty is attributed to unobservable quantities, goodness-of-fit for such models cannot be assessed. 

Aiming at improving the quality of model fit, Balakrishnan and Peng~\cite{Balakrishnan2006} proposed the generalized gamma (GG) frailty model. The GG family of distributions includes exponential, gamma, Weibull etc., all as special members of the family. Therefore, a GG distribution provides a very convenient way to model frailty, as it is possible to choose an appropriate model for the frailty within the GG family for any given data, thus resulting in a better model fit compared to the scenario where only a specific frailty distribution is fitted to the data~\cite{Balakrishnan2006}. 

In this paper, we present a doubly-flexible model for two-component load-sharing systems. In the proposed model, we use the GFB distribution for the baseline of the component lifetimes, and the GG family of distributions to incorporate a shared frailty that captures dependence between the component lifetimes. This model is doubly-flexible, as both for the baseline as well as for the shared random effect, we use general families of distributions, GFB and GG, respectively. Thus, the GFB-GG model structure leads to a two-way general class of models for two-component load-sharing systems. For any given data on a two-component load-sharing system, one can fit the proposed model and then choose suitable distributions for the baseline as well as the shared frailty, within the respective families of distributions used for the baseline and the shared frailty. This way, this GFB-GG model structure provides better fit to two-component load-sharing systems compared to existing models. In summary, among parametric models considered for two-component load-sharing systems in the literature, the proposed model presents a very general case that accommodates load-sharing data with a wide variety of baseline component lifetime distributions and dependence structure. 

The rest of the paper is organized as follows. In Section~\ref{sec:background}, we give brief accounts on the GFB model and the GG frailty model, as preliminaries to further developments. Section~\ref{sec:Model} presents the proposed GFB-GG model for load-sharing data and the fitting method in detail. Results of an elaborate Monte Carlo simulation study, demonstrating the performance of the fitting method and model selection method in this setting, are presented in Section~\ref{sec:simulation}. Numerical examples are presented in Section~\ref{sec:dataana}, wherein a simulation case demonstrates the benefit of using the GFB-GG model structure for a two-component load-sharing system. Analysis of a real dataset is also presented in this section for illustrative purposes. Finally, the paper is concluded in Section~\ref{sec:con} with some remarks.

\section{Background}\label{sec:background}
\subsection{The GFB model}\label{subsec:GFB}
Consider a two-component load-sharing parallel system where components are not replaced or repaired when they fail. Let $C_1$ and $C_2$ denote the two components, with latent lifetimes $T_1$ and $T_2$, respectively. Depending on which of the two components fails first, the lifetime of the other component changes. Allow $T_1$ and $T_2$ to be independent, but not identically distributed. Specifically, assume that $T_1$ and $T_2$ have reliability functions from a PFR class with baseline reliability function $R_B(\cdot; \theta_B)$, and power parameters $\theta_i,$ $i=1, 2,$, respectively. That is, 
\begin{align*}
R_1(t)= P(T_1>t)&=\left[R_B(t, \theta_B)\right]^{\theta_1} \\
\intertext{ and }
R_2(t)= P(T_2>t)&=\left[R_B(t, \theta_B)\right]^{\theta_2}.
\end{align*}                                                           

For the load-sharing setting, we assume that when a component fails, the distribution of the lifetime of the surviving component changes. Here, two scenarios may arise depending on which component fails first: \\
\textbf{Case-1:} $C_1$ fails before $C_2$, i.e., $T_1<T_2$. In this scenario, lifetime of $C_2$ changes from $T_2$ to $T_2^*$ that belongs to a PFR class with baseline reliability $R_B^*(\cdot, \theta_B^*)$ and PFR parameter $\theta_2^*$.\\
\textbf{Case-2:} $C_2$ fails before $C_1$, i.e., $T_1>T_2$. In this scenario, lifetime of $C_1$ changes from $T_1$ to $T_1^*$ that belongs to a PFR class with baseline reliability $R_B^*(\cdot, \theta_B^*)$ and PFR parameter $\theta_1^*$.

Let $(Y_1, Y_2)$ denote the lifetimes of the two components: 
\begin{align*}
\left(Y_1, Y_2\right) = \begin{cases}
\left(T_1^*, T_2\right), & \text{if } {T_1>T_2}\\
\left(T_1, T_2^*\right), & \text{if } {T_1<T_2}.
\end{cases}
\end{align*} 
Let $\lambda_{i0}$ denote the failure rate of the $i$-th component at time $y$ when no component has failed. Then, according to the GFB model,  
\begin{align}
\lambda_{10}(y)&=\theta_1r_B(y, \theta_B),~y>0, \nonumber \\
\lambda_{20}(y)&=\theta_2r_B(y, \theta_B),~y>0, \label{eq:fail-rate-comp}
\end{align}
where $r_B(\cdot, \theta_B)$ denotes the baseline hazard function of the lifetimes of the components before the occurrence of the first failure. Let the $j$-th component fail first. Then, the failure rate of the $i$-th component at time $y$, when the $j$-th component has failed at time $x$, denoted by $\lambda_{ij}(y|x)$, is given by  
\begin{align}
\lambda_{12}(y|x)&=\theta_1^*r_B^*(y, \theta_B^*),~ \text{ if } y>x>0, \nonumber \\
\lambda_{21}(y|x)&=\theta_2^*r_B^*(y, \theta_B^*),~ \text{ if } y>x>0.  \label{eq:fail-rate-comp-1}
\end{align}
where $r_B^*(\cdot, \theta_B^*)$ denotes the baseline hazard function of the lifetimes of the components after the first failure has occurred.

The joint probability density function (PDF) of $(Y_1, Y_2)$ is derived by Franco et al.~\cite{FVK2020} and given by
\begin{align*}
f_{(Y_1,Y_2)}((y_1,y_2))&=\begin{cases}
\theta_1^* \theta_2 r_B^*(y_1, \theta_B^*)r_B(y_2, \theta_B)\left[\frac{R_B^*(y_1, \theta_B^*)}{R_B^*(y_2, \theta_B^*)}\right]^{\theta_1^*}\left[R_B(y_2, \theta_B)\right]^{(\theta_1+\theta_2)}, \text{ if } y_1>y_2>0\\
\theta_1 \theta_2^* r_B(y_1, \theta_B)r_B^*(y_2, \theta_B^*)\left[\frac{R_B^*(y_2, \theta_B^*)}{R_B^*(y_1, \theta_B^*)}\right]^{\theta_2^*}\left[R_B(y_1, \theta_B)\right]^{(\theta_1+\theta_2)}, \text{ if } y_2>y_1>0.
\end{cases}%\label{Joint_pdf}
\end{align*}
This is the GFB distribution by Franco et al.~\cite{FVK2020}, and we denote $$(Y_1,Y_2)\sim GFB(R_B, R_B^*, \theta_1, \theta_2, \theta_1^*, \theta_2^*, \theta_B, \theta_B^*),$$ with PFR parameters $\theta_i, \theta_i^*>0$ for $i=1,2$, and baseline parameters $\theta_B, \theta_B^*$. The GFB family of distributions can be used to generate new bivariate lifetime models by combining different baseline reliability functions $R_B$ and $R_B^*$. In particular, when $R_B=R_B^*$, it resembles the EFB model of Asha et al.~\cite{AKK2016}.

\subsection{The GG distribution}\label{subsec:GG frailty}
The GG distribution, denoted by $GG(\theta, k, \beta)$, has the PDF
\begin{align}
g_Z(z, \theta, k, \beta)= \frac{k}{\theta^{k \beta}\Gamma(\beta)} z^{k \beta-1}e^{-(\frac{z}{\theta})^k}, ~z>0, \label{eq: GG pdf}
\end{align}
with $\theta > 0$ as the scale parameter, and $k, \beta>0$ as shape parameters. The mean and variance, respectively, of the distribution are 
$$E(Z) = \theta\cdot \frac{\displaystyle\Gamma\!\left(\beta+\frac{1}{k}\right)}{\displaystyle\Gamma(\beta)} \quad \text{and} \quad Var(Z) = \theta^2\cdot \left[\frac{\displaystyle\Gamma\!\left(\beta+\frac{2}{k}\right)}{\displaystyle\Gamma(\beta)}-\left\{\frac{\displaystyle\Gamma\!\left(\beta+\frac{1}{k}\right)}{\displaystyle\Gamma(\beta)}\right\}^2\right].$$
%Now, to make the parameters in the stated distribution identifiable, it is usually needed that the mean of the frailty distribution should be one. So,
%\begin{align*}
%\begin{split}
%&E(Z)=1\\
%\implies &\theta\cdot \frac{\Gamma(k+\frac{1}{\beta})}{\Gamma(k)}=1\\
%\implies &\theta=\frac{\Gamma(k)}{\Gamma(k+\frac{1}{\beta})}
%\end{split}
%\end{align*}
%Now, from \eqref{pdf_Z}, we can write the pddf of Z as:
%\begin{align}
%g_Z(z, k, \beta)= \frac{a^k\beta}{\Gamma(k)} z^{k\beta-1}e^{-az^\beta}, ~z>0,
%\end{align}\label{new_pdf_Z}
%where $a=\left(\frac{\Gamma(k+\frac{1}{\beta})}{\Gamma(k)}\right)^\beta >0.$\\

The GG distribution is a family of distributions that includes some well-known distributions such as the exponential, gamma, Weibull etc. as special members of the family, and the lognormal distribution as a limiting case. Table \ref{tab:GG family} gives the choices of parameters for the special members of the GG family of distributions. 
\begin{table}[h]% \scriptsize
	\centering
	\caption{Special members of the GG family of distributions}
	\begin{tabular}{|c|c|}
		\toprule
		Distribution  &  Parameter choice\\
		\midrule
		Exponential & $k=\beta=1$\\
            Gamma & $k=1$ \\
            Weibull & $\beta=1$ \\
            Positive stable with index $\frac{1}{2}$  &  $\beta=\frac{1}{2},~ k=-1$ \\
            Half-normal &  $\beta=\frac{1}{2},~ k=2$ \\
            Lognormal & $\beta \rightarrow 0$ \\
		\bottomrule
	\end{tabular}
		\label{tab:GG family}
\end{table}	
Balakrishnan and Peng~\cite{Balakrishnan2006} advocated for the use of the GG frailty model, highlighting its flexibility derived from its shape parameters. As the PDF of the GG family of distributions takes various forms depending on the parameter values, a GG frailty model is thus capable of describing various types of dependence between the units that share the frailty. For more details on the GG frailty model, refer to Balakrishnan and Peng~\cite{Balakrishnan2006}.

\section{A doubly-flexible model for two-component load-sharing systems}\label{sec:Model}
Suppose that $Y_1$ and $Y_2$ are the lifetimes of the two components of a load-sharing parallel system. Assume that the distribution of $(Y_1, Y_2)$, conditional on the unobserved random factor $Z$, follows the GFB distribution. More specifically, the proposed model can be described in terms of the failure rates of components, that follows from \eqref{eq:fail-rate-comp}, as follows. Conditional on the unobserved random factor $Z$ which we referred to as the shared frailty, the failure rates of the components when both of them are operational, are given by  
\begin{align*}
\lambda_{10}(y|z)&=z\theta_1r_B(y, \theta_B),~y>0,\\
\lambda_{20}(y|z)&=z\theta_2r_B(y, \theta_B),~y>0.
\end{align*}
Now, suppose the $j$-th component fails first at time $y_j$. Then, conditional on the shared frailty $Z$ and the failure time $y_j$, the failure rate of the $i$-th component follows from \eqref{eq:fail-rate-comp-1} as  
\begin{align*}
\lambda_{12}(y_1|y_2,z)&=z\theta_1^*r_B^*(y_1, \theta_B^*),~ \text{ if } y_1>y_2>0,\\
\lambda_{21}(y_2|y_1,z)&=z\theta_2^*r_B^*(y_2, \theta_B^*),~ \text{ if } y_2>y_1>0.
\end{align*}
Correspondingly, the conditional joint PDF of $Y_1$ and $Y_2$ is given by (see Franco et al.~\cite{FVK2020} for futher details)
\begin{align*}
f((y_1,y_2)|z)&=\begin{cases}
z^2\theta_1^* \theta_2 r_B^*(y_1)r_B(y_2)\left[\frac{R_B^*(y_1)}{R_B^*(y_2)}\right]^{z\theta_1^*}\left[R_B(y_2)\right]^{z(\theta_1+\theta_2)}, \text{ if } y_1>y_2>0\\
z^2\theta_1 \theta_2^* r_B(y_1)r_B^*(y_2)\left[\frac{R_B^*(y_2)}{R_B^*(y_1)}\right]^{z\theta_2^*}\left[R_B(y_1)\right]^{z(\theta_1+\theta_2)}, \text{ if } y_2>y_1>0. \label{eq:model-joint-pdf}
\end{cases}
\end{align*}

The shared frailty $Z$ is assumed to follow the GG family of distributions. Thus, the hierarchical representation of the proposed model for component lifetimes of two-component load-sharing system is:
\begin{align*}
  & (Y_1,Y_2) | (Z=z) \sim GFB(R_B, R_B^*, z\theta_1, z\theta_2, z\theta_1^*, z\theta_2^*, \theta_B, \theta_B^*),\\
  & Z \sim GG(\theta, k, \beta).  
\end{align*}
Clearly, the marginal PDF of the two-dimensional lifetime $(Y_1, Y_2)$ can be obtained as 
\begin{align*}
f(y_1,y_2)&=\displaystyle\int_{z=0}^{\infty}f((y_1,y_2)|z) g_Z(z)dz\nonumber\\
&=\begin{cases}
\theta_1^* \theta_2 r_B^*(y_1)r_B(y_2)\displaystyle\int_{z=0}^{\infty}z^2\left[\frac{R_B^*(y_1)}{R_B^*(y_2)}\right]^{z\theta_1^*}\left[R_B(y_2)\right]^{z(\theta_1+\theta_2)}g_Z(z)dz, \text{ if } y_1>y_2>0\\
\theta_1 \theta_2^* r_B(y_1)r_B^*(y_2)\displaystyle\int_{z=0}^{\infty}z^2\left[\frac{R_B^*(y_2)}{R_B^*(y_1)}\right]^{z\theta_2^*}\left[R_B(y_1)\right]^{z(\theta_1+\theta_2)}g_Z(z)dz, \text{ if } y_2>y_1>0.
\end{cases}
\end{align*}

It is a requirement for shared frailty models that the mean of the frailty distribution is fixed at one, for the parameters of the frailty distribution to be identifiable~\cite{Balakrishnan2006}. Using this condition here, we have 
\begin{equation}
E(Z)=1 \implies \theta=\frac{\Gamma(\beta)}{\Gamma(\beta+\frac{1}{k})}. \label{eq:identifiability}
\end{equation}
Using \eqref{eq:identifiability} in \eqref{eq: GG pdf}, the PDF of the frailty distribution becomes
\begin{align*}
g_Z(z, k, \beta)= \frac{ka^\beta}{\Gamma(\beta)} z^{k\beta-1}e^{-az^k}, ~z>0,
\end{align*}
where $a=\left(\frac{\Gamma(\beta+\frac{1}{k})}{\Gamma(\beta)}\right)^k >0.$

The lognormal distribution is a limiting case of the GG distribution when $\beta \rightarrow 0$. For lognormal frailty, after reparameterization using the identifiability condition of setting mean at one, the PDF of the frailty $Z$ becomes
\begin{align*}
g_Z(z, \sigma)= \frac{1}{z\sigma\sqrt(2\pi)}e^{-\frac{(\log(z)+\frac{\sigma^2}{2})^2}{2\sigma^2}}, ~z>0,~\sigma>0.%\label{lognormal-pdf}
\end{align*}

\section{Model fitting method} \label{sec:LikInf}
Consider $n$ two-component load-sharing systems, with observed component lifetimes as $(y_{1i}, y_{2i})$, $i=1,...,n$. The unconditional likelihood function for the proposed model, as presented in Section \ref{sec:Model}, based on the observed data is given by
\begin{align*}
\mathbb{L}(\boldsymbol \Theta)=
\displaystyle\prod_{i=1}^{n}\displaystyle\int_{0}^{\infty}f((y_{1i},y_{2i})|z_i) g_Z(z_i)dz_i,
\end{align*}
where $\boldsymbol \Theta=(\boldsymbol \Theta_1, \boldsymbol \Theta_2)$, with $\boldsymbol \Theta_1 = (\theta_1, \theta_2, \theta_1^*, \theta_2^*, \theta_B, \theta_B^*)$ and $\boldsymbol \Theta_2 = (k, \beta)$, represents the vector of model parameters to be estimated. The corresponding log-likelihood function is
\begin{align*}
\log\mathbb{L}(\boldsymbol \Theta)&=\sum_{i=1}^{n}\displaystyle \log \left( \int_{0}^{\infty}f((y_{1i},y_{2i})|z_i) g_Z(z_i)dz_i \right)\\
&=\sum_{i \in I_1}\log\left({\theta_1^* \theta_2 r_B^*(y_{1i})r_B(y_{2i})}\right)+\sum_{i \in I_2}\log\left({\theta_1 \theta_2^* r_B(y_{1i})r_B^*(y_{2i})}\right)\\&~~+\sum_{i \in I_1}\log\left(\displaystyle\int_{0}^{\infty}z_i^2\left[\frac{R_B^*(y_{1i})}{R_B^*(y_{2i})}\right]^{z_i\theta_1^*}\left[R_B(y_{2i})\right]^{z_i(\theta_1+\theta_2)}g_Z(z_i)dz_i\right)\\&~~+\sum_{i \in I_2}\log\left(\displaystyle\int_{0}^{\infty}z_i^2\left[\frac{R_B^*(y_{2i})}{R_B^*(y_{1i})}\right]^{z_i\theta_2^*}\left[R_B(y_{1i})\right]^{z_i(\theta_1+\theta_2)}g_Z(z_i)dz_i\right),
\end{align*}
where 
\begin{align*}
I_1=\left\{i: y_{1i}>y_{2i}\right\}, ~ |I_1|=n_1 \text{ and }
I_2=\left\{i: y_{1i}<y_{2i}\right\}, ~ |I_2|=n_2.
% I&=I_1\cup I_2=\{1, 2, \ldots,n\}, ~ |I|=n_1+n_2=n.
\end{align*}
Clearly, finding the maximum likelihood estimates (MLEs) of the model parameters is a challenging task as it involves an eight-dimensional numerical optimization of a function without an explicit expression. In such cases, the Expectation-Maximization (EM) algorithm is a very useful tool that offers a reliable method for estimation~\cite{dempster1977}. The EM algorithm is a general approach for estimation when the available data can be perceived incomplete. The Expectation (E) and Maximization (M) steps of the algorithm proceed iteratively. First, a pseudo-complete likelihood function is constructed considering the hypothetical situation of full availability of data; then, conditional expectation of the pseudo-complete likelihood is calculated and maximized with respect to the parameters, putting the entire process in an iterative format. Although a powerful tool, implementation of the traditional EM algorithm can be quite a challenging task sometimes, as it requires calculation of expectations of the pseudo-complete likelihood function conditional on the observed data. Both analytical and numerical complications, such as handling intractable integrals and saddle point traps for successive estimates, are common for complex models. Various types of EM algorithms, such as the Gradient EM, the Monte Carlo EM, or the Stochastic EM, have been proposed to mitigate these challenges, and are applied to practical scenarios. For details on the EM algorithm and its many variants, see McLachlan and Krishnan~\cite{McLachlan2008}. Here we apply an EM-type algorithm wherein we approximate the required conditional expectations by their Monte Carlo estimates, generating samples from suitable conditional distributions.

\subsection{Implementation of the EM-type algorithm}\label{subsec:EM}

%In the E-step, one needs to compute the conditional expectation of the log-likelihood with respect to missing data given the observed data. Then, in the M-step, the maximizer of this expected log-likelihood is computed.  Let $\boldsymbol{\theta}$ be the vector of unknown parameters. Denote the observed part of complete data $\boldsymbol{x}= (x_1 , x_2 ,\ldots, x_n )$ by $\boldsymbol{y}= (y_1 , y_2 ,\ldots, y_m )$ and the missing part by $\boldsymbol{z} = (z_{m+1} , z_{m+2}, \ldots, z_n )$. The complete-data likelihood is
%\begin{align*}
%L_c(\boldsymbol{\theta}|\boldsymbol{y}, \boldsymbol{z})= \prod_{i=1}^{m}f(y_i|\boldsymbol{\theta})\prod_{i={m+1}}^{n}f(z_i|\boldsymbol{\theta})= \prod_{i=1}^{n}f(x_i).
%\end{align*}
%Denote the estimate of $\boldsymbol{\theta}$ at the $s$-th iteration of EM sequences by $\boldsymbol{\theta}_s$. Then, the EM algorithm can be implemented as follow:
%\begin{itemize}
%	\item E-step: Compute
%	\begin{align*}
%	Q(\boldsymbol{\theta}|\boldsymbol{\theta}_s)=E\left[\log
%	L_c(\boldsymbol{\theta}|\boldsymbol{y},\boldsymbol{z})\right]=\int \log
%	%L_c(\boldsymbol{\theta}|\boldsymbol{y},\boldsymbol{z})f(\boldsymbol{z}|\boldsymbol{y},
%	\boldsymbol{\theta}_s).
%	\end{align*}
%	\item M-step: Find $\boldsymbol{\theta}_{s+1}$, which maximizes
%	$Q(\boldsymbol{\theta}|\boldsymbol{\theta}_s)$ in $\boldsymbol{\theta}$.
%\end{itemize}
%For more details, the interested readers are referred to McLachlan et al.~\cite{McLachlan2008} and the citations therein.

For a sample of $n$ two-component load-sharing systems, the hypothetical complete data would consist of the observed data $\{(y_{1i}, y_{2i}), i=1,2,\ldots,n\}$ along with shared frailty $\{z_i:i=1,2,\ldots, n\}$. In reality, of course, $\{z_i:i=1,2,\ldots, n\}$ are not observed, and thus can be considered as the missing data. Therefore, augmenting $\{z_i:i=1,2,\ldots, n\}$ with the observed data, the pseudo-complete data would be of the form 
$$\{(y_{1i}, y_{2i}, z_i), i=1,2,\ldots,n\}.$$
%$$\begin{bmatrix}
%y_{11} & 	y_{22} & z_{1}\\
%y_{12} & 	y_{22} & z_{2}\\
%y_{13} & 	y_{23} & z_{3}\\
%\vdots &   \vdots  & \vdots\\
%y_{1n} & 	y_{2n} & z_{n}\\
%\end{bmatrix}$$
Based on the pseudo-complete data, the pseudo-complete likelihood function for the proposed model is 
\begin{align*}
\mathbb{L}_c(\boldsymbol \Theta)=\prod_{i=1}^{n}f((y_{1i},y_{2i})|z_i)g_Z(z_i)
\end{align*}
with the corresponding pseudo-complete log-likelihood function
\begin{align}
\log\mathbb{L}_c(\boldsymbol \Theta)&=\sum_{i=1}^{n}\log(f((y_{1i},y_{2i})|z_i)g_Z(z_i)) \nonumber \\
&=\sum_{i=1}^{n}\log(f((y_{1i},y_{2i})|z_i) +\sum_{i=1}^{n}\log(g_Z(z_i))) \nonumber \\
%&=2\sum_{i=1}^{n}\log(z_i)+n_1\log(\theta_1^*\theta_2)+n_2\log(\theta_1\theta_2^*)+\sum_{i\in I_1}\log(r_B^*(y_{1i}))+\sum_{i\in I_1}\log(r_B(y_{2i}))\nonumber \\
&~~~+\theta_1^*\sum_{i \in I_1}z_i\log\left(\frac{R_B^*(y_{1i})}{R_B^*(y_{2i})}\right)+(\theta_1+\theta_2)\sum_{i \in I_1}z_i\log(R_B(y_{2i}))+\sum_{i\in I_2}\log(r_B(y_{1i})) \nonumber \\
&~~~+\sum_{i\in I_2}\log(r_B^*(y_{2i}))+\theta_2^*\sum_{i \in I_2}z_i\log\left(\frac{R_B^*(y_{2i})}{R_B^*(y_{1i})}\right)+(\theta_1+\theta_2)\sum_{i \in I_2}z_i\log(R_B(y_{1i})) \nonumber \\
&~~~+n\log\left(\frac{ka^\beta}{\Gamma(\beta)}\right)+(k\beta-1)\sum_{i=1}^{n}\log(z_i)-a\sum_{i=1}^{n}z_i^k.\label{eq:pc-exp-lik}
\end{align}
Here, the expected log-likelihood function is
\begin{align*}
E\left(\log\mathbb{L}_c(\boldsymbol \Theta)\right)=\displaystyle\int_{z=0}^{\infty}\log\mathbb{L}_c(\boldsymbol \Theta)g_Z(z)dz.
\end{align*}

The conditional expectations to be calculated here are $E(Z|(Y_1, Y_2))$, $E(\log(Z)|(Y_1, Y_2)),$ and $E(Z^k|(Y_1, Y_2))$, which cannot be analytically evaluated. Therefore, we use a method wherein we generate samples from the conditional distribution of $Z|(Y_1, Y_2)$, and using the samples, calculate Monte Carlo estimates of the conditional expectations, as described below.   

\subsubsection*{Evaluation of the conditional expectations} 
We generate $N$ samples of $\{z_{i}: i=1, 2, \ldots, N\}$, from the conditional distribution of $Z|(Y_1, Y_2)$. The conditional PDF $f(z|y_1,y_2)$ is given by 
\begin{align*}
f_{Z|(Y_1,Y_2)}(z|(y_1,y_2))&=\begin{cases}
B_1z^{k\beta+1}e^{-az^k-b_1z}, & \text{ if } y_1>y_2>0\\
B_2z^{k\beta+1}e^{-az^k-b_2z}, & \text{ if } y_2>y_1>0,
\end{cases}
\end{align*}
where
\begin{align*}
B_1&= \frac{ka^\beta}{D_1},\\
B_2&= \frac{ka^\beta}{D_2},\\
A_1&=\left[\frac{R_B^*(y_1)}{R_B^*(y_2)}\right]^{\theta_1^*}\left[R_B(y_2)\right]^{(\theta_1+\theta_2)},\\
A_2&=\left[\frac{R_B^*(y_2)}{R_B^*(y_1)}\right]^{\theta_2^*}\left[R_B(y_1)\right]^{(\theta_1+\theta_2)},\\
D_1&=\int_{0}^{\infty}z^2A_1^z\cdot g_Z(z)dz,\\
D_2&=\int_{0}^{\infty}z^2A_2^z\cdot g_Z(z)dz,\\
b_1&= -\log(A_1)>0 \text{ and } b_2= -\log(A_2)>0.
\end{align*}
For $a,~ \beta,~ z>0$, we have $e^{-az^k}\leq 1$. Therefore,
\begin{align*}
f(z|y_1,y_2)&\leq \begin{cases}
B_1z^{k\beta+1}e^{-b_1z}, \text{ if } y_1>y_2>0\\
B_2z^{k\beta+1}e^{-b_2z}, \text{ if } y_2>y_1>0
\end{cases}\\
&\leq \begin{cases}
C_1\cdot \frac{b_1^{k\beta+2}}{\Gamma(k\beta+2)}z^{(k\beta+2)-1}e^{-b_1z}, \text{ if } y_1>y_2>0\\
C_2\cdot \frac{b_2^{k\beta+2}}{\Gamma(k\beta+2)}z^{(k\beta+2)-1}e^{-b_2z}, \text{ if } y_2>y_1>0
\end{cases}\\
&\leq \begin{cases}
C_1\cdot h_1(z), \text{ if } y_1>y_2>0\\
C_2\cdot h_2(z), \text{ if } y_2>y_1>0,
\end{cases}
\end{align*}
where $C_\ell=\frac{B_\ell\Gamma(k\beta+2)}{b_\ell^{k\beta+2}}$ and $h_\ell(z)$ is the PDF of a Gamma random variable with shape parameter $(k\beta+2)$ and rate parameter $b_\ell$ for $\ell=1,~2$. Finally, using the acceptance-rejection method, we can generate samples ${z_1, z_2, \ldots, z_N}$ from $f_{Z|(Y_1,Y_2)}$. Using the generated sample, Monte Carlo estimate of expectation of any function $h(\cdot)$ of $Z$ can be approximated as
$$ 
E(h(Z))=\frac{1}{N}\sum_{i=1}^{N}h(z_i).
$$

\subsubsection*{Maximization step} 
In the M-step, it is convenient to obtain expressions for the PFR parameters $\theta_1, \theta_2, \theta_1^*$, and $\theta_2^*$, by differentiating \eqref{eq:pc-exp-lik} with respect to the parameters, and equating them to zero: 
\begin{align}
\hat{\theta_1}&=\frac{-n_2}{\sum_{i \in I_1}E(Z_i|(Y_{1i},Y_{2i}))\log(R_B(y_{2i}))+\sum_{i \in I_2}E(Z_i|(Y_{1i},Y_{2i}))\log(R_B(y_{1i}))}, \nonumber \\
\hat{\theta_2}&=\frac{-n_1}{\sum_{i \in I_1}E(Z_i|(Y_{1i},Y_{2i}))\log(R_B(y_{2i}))+\sum_{i \in I_2}E(Z_i|(Y_{1i},Y_{2i}))\log(R_B(y_{1i}))}, \nonumber \\
\hat{\theta_1^*}&=\frac{-n_1}{\sum_{i \in I_1}E(Z_i|(Y_{1i},Y_{2i}))\log\left(\frac{R_B(y_{1i})}{R_B(y_{2i})}\right)},\nonumber \\
\hat{\theta_2^*}&=\frac{-n_2}{\sum_{i \in I_2}E(Z_i|(Y_{1i},Y_{2i}))\log\left(\frac{R_B(y_{2i})}{R_B(y_{1i})}\right)}. \label{eq:PFR-ests}
\end{align} 
Note that, $\hat{\theta_1}, ~\hat{\theta_2}, ~\hat{\theta_1}^*,~ \hat{\theta_2}^*$ are functions of $\theta_B$ and $\theta_B^*$.

Using \eqref{eq:PFR-ests} in \eqref{eq:pc-exp-lik} gives the profile log-likelihood function (up to an additive constant) of $\tilde{\boldsymbol \Theta_1}=(\theta_B, \theta_B^*)$ and $\boldsymbol \Theta_2$ as
\begin{align*}
p(\tilde{\boldsymbol \Theta_1}, \boldsymbol \Theta_2)&=
n_1\log(\hat{\theta_1^*}\hat{\theta_2})+\sum_{i\in I_1}\log(r_B^*(y_{1i}))+\sum_{i\in I_1}\log(r_B(y_{2i}))\\%-2n\\
&~~~+n_2\log(\hat{\theta_1}\hat{\theta_2^*})+\sum_{i\in I_2}\log(r_B(y_{1i}))
+\sum_{i\in I_2}\log(r_B^*(y_{2i}))\\
&~~~+n\log\left(\frac{ka^\beta}{\Gamma(\beta)}\right)+(k\beta-1)\sum_{i=1}^{n}E(\log(Z_i|(Y_{1i},Y_{2i})))-a\sum_{i=1}^{n}E((Z_i|(Y_{1i},Y_{2i}))^k)\\
&=H_1(\tilde{\boldsymbol \Theta_1}) + H_2(\boldsymbol \Theta_2),
\end{align*}
where
\begin{align*}
H_1(\tilde{\boldsymbol \Theta_1})& = n_1\log(\hat{\theta_1^*}\hat{\theta_2})+\sum_{i\in I_1}\log(r_B^*(y_{1i}))+\sum_{i\in I_1}\log(r_B(y_{2i}))\\%-2n\\
&~~~+n_2\log(\hat{\theta_1}\hat{\theta_2^*})+\sum_{i\in I_2}\log(r_B(y_{1i}))
+\sum_{i\in I_2}\log(r_B^*(y_{2i})),\\
H_2(\boldsymbol \Theta_2)& = n\log\left(\frac{ka^\beta}{\Gamma(\beta)}\right)+(k\beta-1)\sum_{i=1}^{n}E(\log(Z_i|(Y_{1i},Y_{2i})))-a\sum_{i=1}^{n}E((Z_i|(Y_{1i},Y_{2i}))^k).
\end{align*}
Maximization of $p(\tilde{\boldsymbol \Theta_1}, \boldsymbol \Theta_2)$ is relatively simpler, as it is the sum of two two-dimensional functions. Therefore, $H_1(\tilde{\boldsymbol \Theta_1})$ and $H_2(\boldsymbol \Theta_2)$ can be separately maximized to obtain MLEs $\hat{\theta_B}, \hat{\theta_B^*}$, and $\hat{k}, \hat{\beta}$, respectively, using which in \eqref{eq:PFR-ests} finally gives the MLEs $\hat{\theta_1}, \hat{\theta_2}, \hat{\theta_i^*}$, and $\hat{\theta_2^*}$.

\subsection{Model selection}\label{subsec:modelselection}
For a given two-component load-sharing data, it is of prime importance to select the best model within the GFB-GG structure, i.e., the best combination of the baseline and frailty distributions within the GFB and GG families, respectively. Clearly, there is a large number of candidates for the best baseline-frailty combination within the GFB-GG model structure. When all the candidate models are fitted to the given data, the best model can be selected with the help of one of the standard model selection criteria. 

Consider model $M$ with any particular combination of baseline and frailty distribution within the GFB-GG structure, involving $d_M$ number of parameters to be estimated. Let $n$ denote the size of the available sample for model fitting. Suppose ${\displaystyle{\hat{L}_M}}$ is the maximized value of the likelihood function for model $M$. Then, the Akaike's information criterion (AIC), Bayesian information criterion (BIC), finite sample corrected Akaike's information criterion (AICc), and Bridge criterion (BC) for the model $M$ are the following:
\begin{align*}
\displaystyle \mathrm {AIC}_M \,&=\,2d_M-2\ln({\widehat {L}_M}),\\
\displaystyle \mathrm {BIC}_M \,&=\,d_M\cdot\log(n)-2\ln({\widehat {L}_M}),\\
\displaystyle \mathrm {AICc}_M \,&=\displaystyle \mathrm {AIC} + \frac{2d_M(d_M+1)}{n-d_M-1},\\
\displaystyle \mathrm {BC}_M \,&=\,n^{2/3} \sum_{m=1}^{d_M}\frac{1}{k}-2\ln({\widehat {L}_M}).
\end{align*}
Among all the candidate models, the best model for the given data is the one with the minimum AIC (or BIC, AICc, BC) value. Note that BC is developed to combine the strengths of AIC and BIC. In the parametric scenario, BC attains the properties of BIC, and in the nonparametric scenario, it attains the same as AIC adaptively. For details on statistical model selection, see Claeskens~\cite{Claeskens2016}.

\section{Simulation study}\label{sec:simulation}
The Monte Carlo simulations conducted in this study serve two main purposes: 
\begin{enumerate}
    \item It demonstrates the performance of the model fitting method as described in Section~\ref{sec:LikInf}, with respect to fitting characteristics such as average estimates (AE), mean squared error (MSE); 
    \item It examines, for a given two-component load-sharing data, the performance of different model selection criteria for choosing the best model with a baseline-frailty combination within the GFB-GG model structure. 
\end{enumerate}
Within the GFB-GG model structure, we consider 27 different combinations of baseline and frailty distributions for the simulations; details of these combinations are presented in Table~\ref{tab:modelname}. 

\subsection{Demonstration of the model fitting method} \label{subsec:sim-evaluation}
For each of the 27 baseline-frailty combinations, the broad steps as listed below are followed, based on 500 Monte Carlo runs: 
\begin{enumerate}
    \item Generate two-component load-sharing data with a specific set of model parameters and simulation parameters (where all parameters are chosen without any loss of generality); 
    \item Estimate the model parameters by using the fitting method of Section \ref{sec:LikInf}, assuming the parent model (i.e., the baseline-frailty combination used for data generation) as the true model;
    \item Compute AE and MSE for the parameter estimates; 
    \item For each parameter, construct confidence intervals by using Louis' missing information principle~\cite{Louis1982}, and compute coverage percentages (CP) and average lengths (AL) of the confidence intervals. 
\end{enumerate}

The sample size used for simulations is 100. The results are furnished in Tables \ref{tab:expfrailty}, \ref{tab:gammafrailty}, and \ref{tab:weibullfrailty}, for different frailty distributions. It is clear that the fitting method works quite well for the proposed model, as the AE, MSE, CP, and AL for all the model parameters are satisfactory. Clearly, it can be expected that for larger sample sizes, the performance of the fitting method would improve.   

\begin{table}[p]
	\caption{Model indicators according to their frailty distributions and baseline distributions before and after 1st failure}
	\vspace{0.2cm}
	\label{tab:modelname}
	%\scriptsize
	%\tabcolsep7pt
	\centering
	\begin{tabular}{|c|c|c|c|}
		\hline
		Model Name & Frailty distribution & Baseline before failure & Baseline after failure\\
		\hline
		M1 & Exp(1) & Exp(1) & Exp(1)\\
		\hline
		M2 & Exp(1) & Exp(1) & Weibull($\theta_B^*$, 1)\\
		\hline
		M3 & Exp(1) & Exp(1) & Gamma($\theta_B^*$, 1)\\
		\hline
		M4 & Exp(1) & Weibull($\theta_B$, 1) & Exp(1)\\
		\hline
		M5 & Exp(1) & Weibull($\theta_B$, 1) & Weibull($\theta_B^*$, 1)\\
		\hline
		M6 & Exp(1) & Weibull($\theta_B$, 1) & Gamma($\theta_B^*$, 1)\\
		\hline
		M7 & Exp(1) & Gamma($\theta_B$, 1) & Exp(1)\\
		\hline
		M8 & Exp(1) & Gamma($\theta_B$, 1) & Weibull($\theta_B^*$, 1)\\
		\hline
		M9 & Exp(1) & Gamma($\theta_B$, 1) & Gamma($\theta_B^*$, 1)\\
		\hline
		M10 & Weibull(shape=$k$, scale=$\frac{1}{\Gamma(1+\frac{1}{k})})$ & Exp(1) & Exp(1)\\
		\hline
		M11 & Weibull(shape=$k$, scale=$\frac{1}{\Gamma(1+\frac{1}{k})})$ & Exp(1) & Weibull($\theta_B^*$, 1)\\
		\hline
		M12 & Weibull(shape=$k$, scale=$\frac{1}{\Gamma(1+\frac{1}{k})})$ & Exp(1) & Gamma($\theta_B^*$, 1)\\
		\hline
		M13 & Weibull(shape=$k$, scale=$\frac{1}{\Gamma(1+\frac{1}{k})})$ & Weibull($\theta_B$, 1) & Exp(1)\\
		\hline
		M14 & Weibull(shape=$k$, scale=$\frac{1}{\Gamma(1+\frac{1}{k})})$ & Weibull($\theta_B$, 1) & Weibull($\theta_B^*$, 1)\\
		\hline
		M15 & Weibull(shape=$k$, scale=$\frac{1}{\Gamma(1+\frac{1}{k})})$ & Weibull($\theta_B$, 1) & Gamma($\theta_B^*$, 1)\\
		\hline
		M16 & Weibull(shape=$k$, scale=$\frac{1}{\Gamma(1+\frac{1}{k})})$ & Gamma($\theta_B$, 1) & Exp(1)\\
		\hline
		M17 & Weibull(shape=$k$, scale=$\frac{1}{\Gamma(1+\frac{1}{k})})$ & Gamma($\theta_B$, 1) & Weibull($\theta_B^*$, 1)\\
		\hline
		M18 & Weibull(shape=$k$, scale=$\frac{1}{\Gamma(1+\frac{1}{k})})$ & Gamma($\theta_B$, 1) & Gamma($\theta_B^*$, 1)\\
		\hline
		M19 & Gamma(shape=$\beta$, scale=$\frac{1}{\beta})$ & Exp(1) & Exp(1)\\
		\hline
		M20 & Gamma(shape=$\beta$, scale=$\frac{1}{\beta})$ & Exp(1) & Weibull($\theta_B^*$, 1)\\
		\hline
		M21 & Gamma(shape=$\beta$, scale=$\frac{1}{\beta})$ & Exp(1) & Gamma($\theta_B^*$, 1)\\
		\hline
		M22 & Gamma(shape=$\beta$, scale=$\frac{1}{\beta})$ & Weibull($\theta_B$, 1) & Exp(1)\\
		\hline
		M23 & Gamma(shape=$\beta$, scale=$\frac{1}{\beta})$ & Weibull($\theta_B$, 1) & Weibull($\theta_B^*$, 1)\\
		\hline
		M24 & Gamma(shape=$\beta$, scale=$\frac{1}{\beta})$ & Weibull($\theta_B$, 1) & Gamma($\theta_B^*$, 1)\\
		\hline
		M25 & Gamma(shape=$\beta$, scale=$\frac{1}{\beta})$ & Gamma($\theta_B$, 1) & Exp(1)\\
		\hline
		M26 & Gamma(shape=$\beta$, scale=$\frac{1}{\beta})$ & Gamma($\theta_B$, 1) & Weibull($\theta_B^*$, 1)\\
		\hline
		M27 & Gamma(shape=$\beta$, scale=$\frac{1}{\beta})$ & Gamma($\theta_B$, 1) & Gamma($\theta_B^*$, 1)\\
		\hline
	\end{tabular}
	%\label{tab:modelname}
\end{table}

\begin{table}[p] \scriptsize
	\centering
	\caption{AE, MSE, AL, CP of the parameters based on 500 simulated samples for size $n=100$, according to load-sharing systems from different models with GFB parameters $\Theta_1=(\theta_1, \theta_2, \theta_1^*, \theta_2^*, \theta_B, \theta_B^*)$ having Exponential(rate=1) frailty distribution.}
	\begin{tabular}{c|c|ccccccc}
		\toprule
		Model  &  Parameters & $\theta_1$  & $\theta_2$  & $\theta_1^*$ & $\theta_2^*$& $\theta_B $ & $\theta_B^*$ &\\
		\cmidrule(lr){2-9}	
		Number & True Value & 0.3 & 0.4 & 0.5 & 1.0 & 2.0 & 1.5 & \\		
		
		\midrule
		
		&AE & 0.306 & 0.409 & 0.514 & 1.035 & - & -& \\
		\cmidrule(lr){2-9}
		M1&MSE & 0.004 & 0.006 & 0.011 & 0.058&- &- &\\
		\cmidrule(lr){2-9}
		&AL & 0.243 & 0.299 & 0.414 & 0.949 & -& -& \\
		\cmidrule(lr){2-9}
		& CP & 95.8 & 94.0 & 94.6 & 95.8 &- &- & \\
		
		\midrule
		
		& AE & 0.307 & 0.411 & 0.515 & 1.033 & - & 1.510 & \\
		\cmidrule(lr){2-9}
		M2  & MSE & 0.004 & 0.006 & 0.018 & 0.077 &-  & 0.012 & \\
		\cmidrule(lr){2-9}
		&AL & 0.253 & 0.313 & 0.521 & 1.096 &  -& 0.431 & \\
		\cmidrule(lr){2-9}
		& CP & 95.8 & 94.6 & 94.6 & 95.0 & -& 95.2 & \\
		
		\midrule
		
		&AE & 0.308 &0.411  & 0.535 & 1.084 & - & 1.553 & \\
		\cmidrule(lr){2-9}
		M3&MSE& 0.004 & 0.006 & 0.021 & 0.109 &-  & 0.148 & \\
		\cmidrule(lr){2-9}
		&AL & 0.250 & 0.309 & 0.540 & 1.271 & - & 1.405 & \\
		\cmidrule(lr){2-9}
		& CP & 95.8 & 94.2 & 95.0 & 95.8 & -& 92.8 & \\
		
		\midrule
		
		&AE & 0.305 & 0.407 & 0.515 & 1.037 & 2.021&- & \\
		\cmidrule(lr){2-9}
		M4&MSE & 0.004 & 0.006 & 0.011 & 0.058 & 0.022 &- & \\
		\cmidrule(lr){2-9}
		&AL & 0.249 & 0.308 & 0.419 & 0.960 & 0.570 & -& \\
		\cmidrule(lr){2-9}
		& CP & 95.6 & 94.8 & 95.0 & 95.6 & 93.2 & -& \\
		
		\midrule
		
		&AE & 0.306 & 0.409 & 0.510 & 1.025 & 2.029 & 1.536 & \\
		\cmidrule(lr){2-9}
		M5&MSE & 0.004 & 0.007 & 0.025 & 0.098 & 0.029 & 0.037 & \\
		\cmidrule(lr){2-9}
		&AL &0.256 &0.318 &0.601 &1.197 &0.640 &0.724 & \\
		\cmidrule(lr){2-9}
		& CP & 96.0& 95.4& 92.8& 93.4& 94.0& 93.4& \\
		
		\midrule
		
		&AE &0.306 &0.409 &0.550 &1.128 &2.024 &1.577 & \\
		\cmidrule(lr){2-9}
		M6&MSE & 0.004 & 0.007 & 0.029& 0.178 & 0.027& 0.285& \\
		\cmidrule(lr){2-9}
		&AL & 0.254 & 0.315 & 0.643 & 1.543& 0.596 & 1.875& \\
		\cmidrule(lr){2-9}
		& CP & 96.0 & 94.8 & 94.8& 95.8& 93.4 & 92.8& \\
		
		\midrule
		
		&AE & 0.312& 0.416& 0.512& 1.031& 2.021&- & \\
		\cmidrule(lr){2-9}
		M7&MSE & 0.007 &0.012 &0.011 &0.058 &0.059 &- & \\
		\cmidrule(lr){2-9}
		&AL & 0.314 &0.401 &0.415 & 0.945 &0.963 &- & \\
		\cmidrule(lr){2-9}
		& CP &94.0 &93.6 &94.6 &95.2 &94.2 &- & \\
		
		\midrule
		
		&AE & 0.320 &0.427 &0.505 &1.013 &2.037 &1.532 & \\
		\cmidrule(lr){2-9}
		M8&MSE & 0.010& 0.015& 0.030&0.120 &0.069 &0.023 & \\
		\cmidrule(lr){2-9}
		&AL & 0.360 &0.463 &0.636 &1.302 &1.029 &0.551 & \\
		\cmidrule(lr){2-9}
		& CP & 92.8 &93.4 &90.6 &91.4 &94.4 &93.4 & \\
		
		\midrule
		
		&AE &0.316 & 0.422 &0.537 &1.093 &2.029 &1.564 & \\
		\cmidrule(lr){2-9}
		M9&MSE & 0.009 & 0.014 &0.024 &0.133 &0.065 &0.413 & \\
		\cmidrule(lr){2-9}
		&AL & 0.339 & 0.433 & 0.588 & 1.358 & 0.990 & 2.200 & \\
		\cmidrule(lr){2-9}
		& CP & 92.4 & 93.8 & 94.4 & 94.8 & 94.0 & 94.2 & \\
		
		\bottomrule
	\end{tabular}
	\label{tab:expfrailty}
\end{table}

\begin{table}[p] \scriptsize
	\centering
	\caption{AE, MSE, AL, CP of the parameters based on 500 simulated samples for size $n=100$, according to load-sharing systems from different models with GFB parameters $\Theta_1=(\theta_1, \theta_2, \theta_1^*, \theta_2^*, \theta_B, \theta_B^*)$ having Weibull(shape=$k$, scale=$\frac{1}{\Gamma(1+\frac{1}{k})}$) frailty distribution with $\Theta_2=k$.}
	\begin{tabular}{c|c|ccccccc}
		\toprule
		Model  &  Parameters & $\theta_1$  & $\theta_2$  & $\theta_1^*$ & $\theta_2^*$& $\theta_B $ & $\theta_B^*$ & $k$\\
		\cmidrule(lr){2-9}	
		Number & True Value & 0.3 & 0.4 & 0.5 & 1.0 & 2.0 & 1.5 & 1.5\\		
		
		\midrule
		
		&AE & 0.304 & 0.405 &0.498  &1.041  & - & -& 1.563 \\
		\cmidrule(lr){2-9}
		M10&MSE & 0.004 & 0.005 & 0.008 & 0.047 &- &- &0.065\\
		\cmidrule(lr){2-9}
		&AL & 0.229 & 0.280 & 0.371 & 0.877 & -& -&0.981 \\
		\cmidrule(lr){2-9}
		& CP & 92.6 & 94.2 & 93.8 & 95.2 &- &- & 95.8\\
		
		\midrule
		
		& AE & 0.307 & 0.409 & 0.511 & 1.059 & - & 1.504 &1.598 \\
		\cmidrule(lr){2-9}
		M11 & MSE & 0.005 & 0.007 & 0.019 & 0.080 &-  & 0.026 & 0.224\\
		\cmidrule(lr){2-9}
		&AL & 0.274 & 0.341 & 0.535 & 1.087 &  -& 0.701 & 1.421\\
		\cmidrule(lr){2-9}
		& CP & 93.4 & 93.0 & 93.8 & 95.8 & -& 94.2 & 94.2\\
		
		\midrule
		
		&AE & 0.309 & 0.411 & 0.526 & 1.120 & - & 1.561 & 1.548\\
		\cmidrule(lr){2-9}
		M12 &MSE & 0.004 & 0.006 & 0.021 & 0.152 & - & 0.182 & 0.083\\
		\cmidrule(lr){2-9}
		&AL & 0.227 & 0.341 & 0.664 & 1.646 & - & 1.945 & 1.451\\
		\cmidrule(lr){2-9}
		& CP & 94.4 & 95.8 & 95.0 & 97.4 & - & 95.0 & 95.6\\
		
		\midrule
		
		& AE & 0.303& 0.405& 0.500& 1.047& 2.007&- & 1.568\\
		\cmidrule(lr){2-9}
		M13 & MSE &0.004 &0.005 &0.010 &0.053 &0.030 &- & 0.073\\
		\cmidrule(lr){2-9}
		& AL & 0.236& 0.290& 0.399& 0.946&0.728 & -& 1.218\\
		\cmidrule(lr){2-9}
		& CP &94.4 &94.4 &93.4 &95.0 &95.4 & -& 95.8\\
		
		\midrule
		
		& AE & 0.316&0.416 &0.502 &1.034 &2.064 &1.588 &1.620 \\
		\cmidrule(lr){2-9}
		M14 & MSE &0.006 &0.008 &0.042 &0.137 &0.071 &0.134 &0.524 \\
		\cmidrule(lr){2-9}
		& AL &0.283 &0.352 &0.754 &1.381 &1.007 &1.328 &1.906 \\
		\cmidrule(lr){2-9}
		& CP & 93.4& 92.4& 87.0& 91.2& 89.8& 86.0&84.2 \\
		
		\midrule
		
		& AE & 0.305 & 0.411 & 0.562 & 1.223 & 2.035 & 1.603 & 1.562 \\
		\cmidrule(lr){2-9}
		M15 & MSE & 0.005& 0.006& 0.049& 0.397&0.045 &0.428 &0.152 \\
		\cmidrule(lr){2-9}
		&AL & 0.260&0.325 &0.923 &2.345 &0.880 &2.625 &1.609 \\
		\cmidrule(lr){2-9}
		& CP & 92.2&94.4 &93.0 &94.8 &94.0 & 94.4& 93.8\\
		
		\midrule
		
		&AE & 0.312& 0.416&0.500 &1.046 &2.016 & -& 1.559\\
		\cmidrule(lr){2-9}
		M16&MSE & 0.008& 0.011&0.009 &0.052 &0.067 &- &0.066 \\
		\cmidrule(lr){2-9}
		&AL & 0.340& 0.435& 0.384&0.908 &1.044 &- &1.084 \\
		\cmidrule(lr){2-9}
		& CP &92.2 &95.4 &93.0 &95.6 &95.4 &- &96.2 \\
		
		\midrule
		
		&AE & 0.334&0.441 &0.534 &1.096 &2.036 &1.510 &1.632 \\
		\cmidrule(lr){2-9}
		M17&MSE & 0.024& 0.035& 0.046& 0.205&0.105 &0.057 & 0.242\\
		\cmidrule(lr){2-9}
		&AL & 0.541&0.696 &0.853 &1.699 & 1.413& 1.045& 1.824\\
		\cmidrule(lr){2-9}
		& CP &92.6 &92.4 &92.8 &93.6 &93.6 & 91.0& 91.6\\
		
		\midrule
		
		&AE & 0.326&0.434 &0.550 &1.159 &2.038 &1.598 & 1.552\\
		\cmidrule(lr){2-9}
		M18&MSE & 0.014&0.020 &0.040 &0.245 &0.083 &0.637 &0.114 \\
		\cmidrule(lr){2-9}
		&AL &0.460 &0.596 &0.828 &1.984 &1.255 &0.637 &0.114 \\
		\cmidrule(lr){2-9}
		& CP &92.8 &95.4 & 93.8& 94.4& 96.0&93.0 &93.6 \\
		
		\bottomrule
	\end{tabular}
		\label{tab:weibullfrailty}
\end{table}	

\begin{table}[p] \scriptsize
	\centering
	\caption{AE, MSE, AL, CP of the parameters based on 500 simulated samples for size $n=100$, according to load-sharing systems from different models with GFB parameters $\Theta_1=(\theta_1, \theta_2, \theta_1^*, \theta_2^*, \theta_B, \theta_B^*)$ having Gamma(shape=$\beta$, scale=$\frac{1}{\beta}$) frailty distribution with $\Theta_2=\beta$.}
	\begin{tabular}{c|c|ccccccc}
		\toprule
		Model  &  Parameters & $\theta_1$  & $\theta_2$  & $\theta_1^*$ & $\theta_2^*$& $\theta_B $ & $\theta_B^*$ & $\beta$\\
		\cmidrule(lr){2-9}	
		Number & True Value & 0.3 & 0.4 & 0.5 & 1.0 & 2.0 & 1.5 & 1.5\\		
		
		\midrule
		
		&AE & 0.303 & 0.404 & 0.508 & 1.028 & - & -& 1.632\\
		\cmidrule(lr){2-9}
		M19 &MSE & 0.004 & 0.005 & 0.011 & 0.052 &- &- &0.166\\
		\cmidrule(lr){2-9}
		&AL & 0.233 & 0.287 & 0.392 & 0.900 & -& -& 1.369\\
		\cmidrule(lr){2-9}
		& CP & 93.6 & 95.8 & 90.4 & 95.4 &- &- & 97.0\\
		
		\midrule
		
		&AE & 0.304 & 0.406 & 0.524 & 1.055 & - & 1.492 & 1.661\\
		\cmidrule(lr){2-9}
		M20 & MSE & 0.005 & 0.006 & 0.020 & 0.075 &-  & 0.021 & 0.263 \\
		\cmidrule(lr){2-9}
		&AL & 0.273 & 0.343 & 0.549 & 1.116 &  -& 0.650 & 1.946\\
		\cmidrule(lr){2-9}
		& CP & 92.2 & 94.0 & 94.4 & 95.4 & -& 96.0 & 96.8 \\
		
		\midrule
		
		&AE & 0.303 &0.406  & 0.520 & 1.059 & - & 1.502 & 1.646\\
		\cmidrule(lr){2-9}
		M21 & MSE & 0.004 & 0.006 & 0.022 & 0.111 &-  & 0.159 & 0.228\\
		\cmidrule(lr){2-9}
		&AL & 0.251 & 0.312 & 0.587 & 1.387 & - & 1.668 & 1.661\\
		\cmidrule(lr){2-9}
		& CP & 92.4 & 94.4 & 92.8 & 94.6 & -& 95.8 & 98.0\\
		
		\midrule
		
		&AE & 0.301& 0.403& 0.510& 1.033& 2.021&- & 1.635\\
		\cmidrule(lr){2-9}
		M22 & MSE &0.004 &0.005 &0.012 &0.053 &0.027 &- &0.214 \\
		\cmidrule(lr){2-9}
		&AL & 0.237& 0.293& 0.413& 0.940 &0.682 & -& 1.593\\
		\cmidrule(lr){2-9}
		& CP &93.2 &95.0 &92.2 &95.2 &96.2 & -& 96.4\\
		
		\midrule
		
		&AE & 0.312 & 0.416 & 0.518 & 1.034 & 2.048 & 1.557 & 1.756\\
		\cmidrule(lr){2-9}
		M23 & MSE & 0.005 & 0.008 & 0.037 & 0.122 & 0.065 & 0.108 & 0.852\\
		\cmidrule(lr){2-9}
		&AL &0.291 &0.360 &0.772 &1.408 &1.007 &1.300 &2.921 \\
		\cmidrule(lr){2-9}
		& CP & 94.6& 94.0& 90.4& 92.2& 91.6& 89.4& 91.2\\
		
		\midrule
		
		&AE &0.304 &0.403 &0.564 &1.164 &2.040 &1.558 &1.652 \\
		\cmidrule(lr){2-9}
		M24 & MSE & 0.004& 0.006 & 0.062 & 0.342& 0.041 & 0.419 & 0.425\\
		\cmidrule(lr){2-9}
		&AL & 0.262 & 0.328 & 0.911 & 2.179& 0.873 & 2.614 & 2.452\\
		\cmidrule(lr){2-9}
		& CP & 94.0 & 94.2 & 93.4 & 95.0& 94.8 & 93.8 & 95.4\\
		
		\midrule
		
		&AE & 0.315& 0.422&0.510 &1.032 &2.037 &- &1.628 \\
		\cmidrule(lr){2-9}
		M25 & MSE &0.007 &0.013 & 0.011& 0.053& 0.064& -& 0.193\\
		\cmidrule(lr){2-9}
		&AL & 0.340& 0.437&0.403 & 0.931& 1.032& -&1.500 \\
		\cmidrule(lr){2-9}
		& CP &94.8 &95.8 &92.6 & 96.2& 95.6& -&97.8 \\
		
		\midrule
		
		&AE & 0.329& 0.438& 0.543& 1.097& 2.045& 1.506&1.698 \\
		\cmidrule(lr){2-9}
		M26 & MSE & 0.018&0.032 & 0.055& 0.189& 0.103&0.049 &0.651 \\
		\cmidrule(lr){2-9}
		&AL & 0.516 & 0.661& 0.858& 1.702& 1.362 & 0.949 & 2.494 \\
		\cmidrule(lr){2-9}
		& CP &91.6 & 91.6& 92.8& 93.6 & 93.2 &93.0 & 94.0\\
		
		\midrule
		
		&AE & 0.328	& 0.439	& 0.544	& 1.120 &	2.052 &	1.537	&1.623 \\
		\cmidrule(lr){2-9}
		M27 & MSE &0.015	&0.030&	0.041&	0.203&	0.087&	0.532&	0.245\\
		\cmidrule(lr){2-9}
		&AL &0.444	&0.582	&0.781	&1.792&	1.208&	2.846&	2.052\\
		\cmidrule(lr){2-9}
		& CP &94.6	&94.8	&93.4	&95.4	&95.2&	96.2&	95.6\\
		
		\bottomrule
	\end{tabular}
	\label{tab:gammafrailty}
\end{table}

\subsection{Study of model selection}  \label{subsec:sim-model}
For model selection, in this setting, our interest is to examine if the standard criteria can accurately detect the parent models with various combinations of baseline-frailty distributions from the GFB-GG model structure. As for the model selection criteria, we use the AIC, BIC, AICc, and BC. The steps in which we conduct the model selection study are as follows: 
\begin{enumerate}
    \item A two-component load-sharing data of size $n$ from a parent model is generated, the parent model being one of the models $M1 - M27$, as described in Table \ref{tab:modelname}; 
    \item To the generated data, all the candidate models $M1 - M27$ are fitted; 
    \item Using the estimated model parameters, the model selection criteria are calculated for each of the fitted candidate models;
    \item By any of the model selection criteria, the model with the lowest value is chosen as the best model for the given data among the 27 candidate models. 
\end{enumerate}

\begin{figure}[p]
	\begin{subfigure}{.5\textwidth}
		\centering
		% include the first image
		\includegraphics[width=\linewidth]{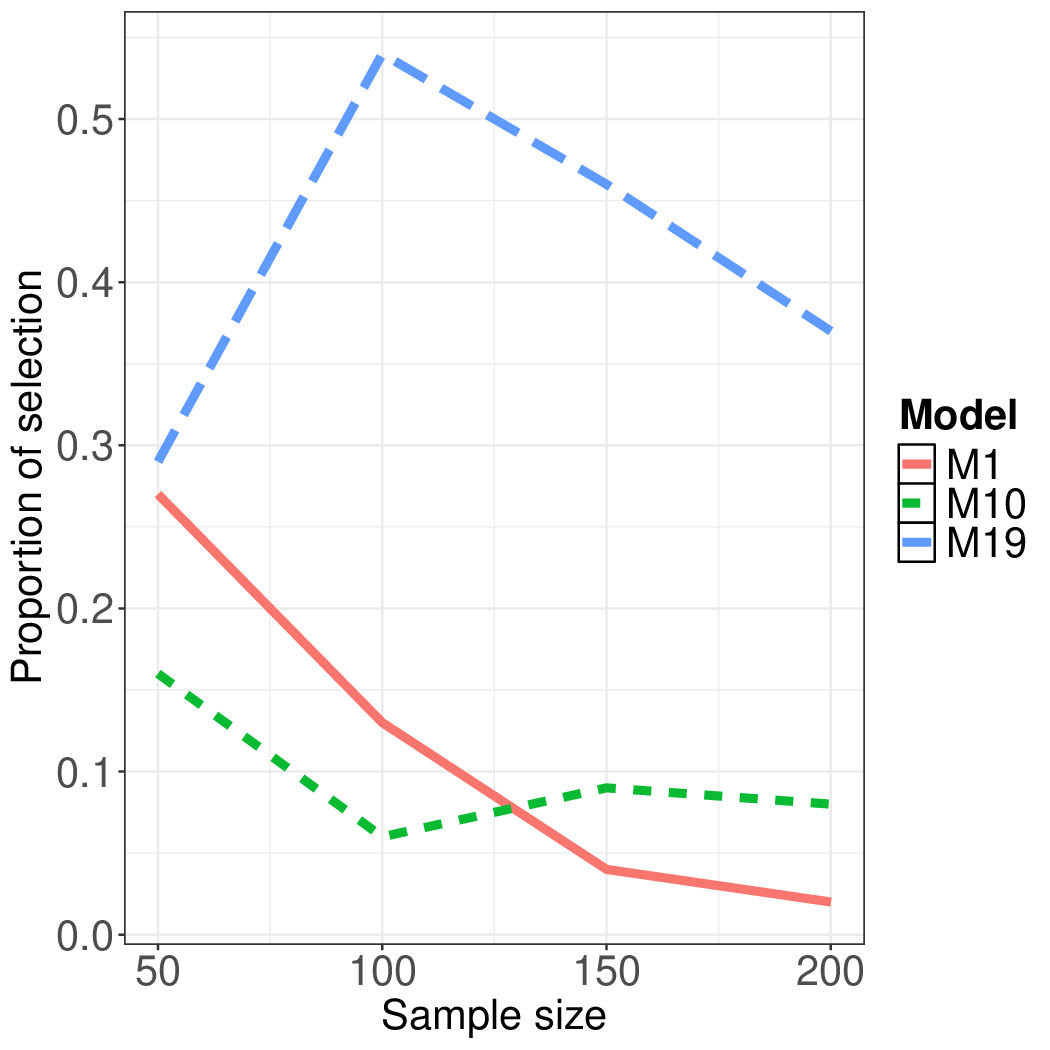}
            \caption{Selection by AIC}
		\label{fig:sub-AIC-M1}
	\end{subfigure}
	\begin{subfigure}{.5\textwidth}
		\centering
		% include the second image
		\includegraphics[width=\linewidth]{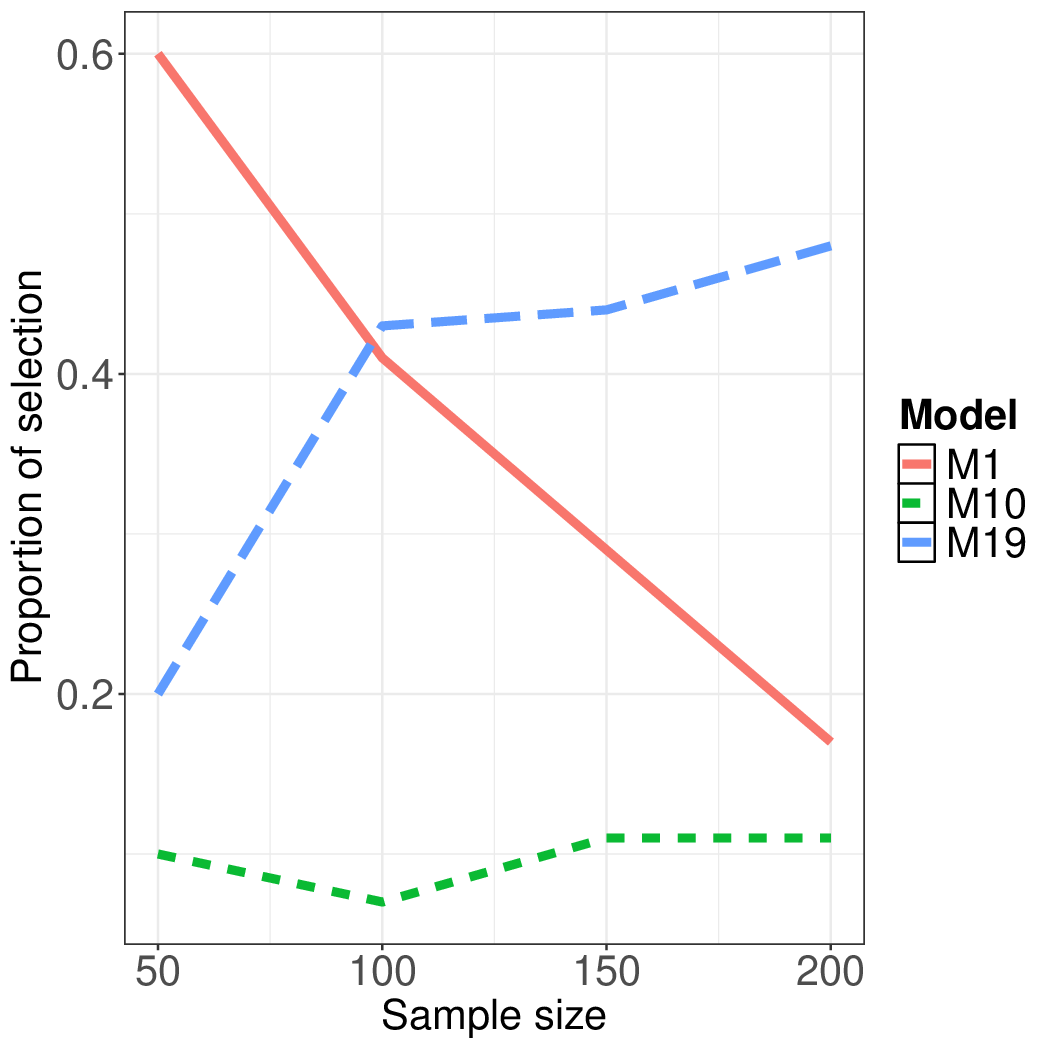}  
		\caption{Selection by BIC}
		\label{fig:sub-BIC-M1}
	\end{subfigure}
	\begin{subfigure}{.5\textwidth}
		\centering
		% include the third image
		\includegraphics[width=\linewidth]{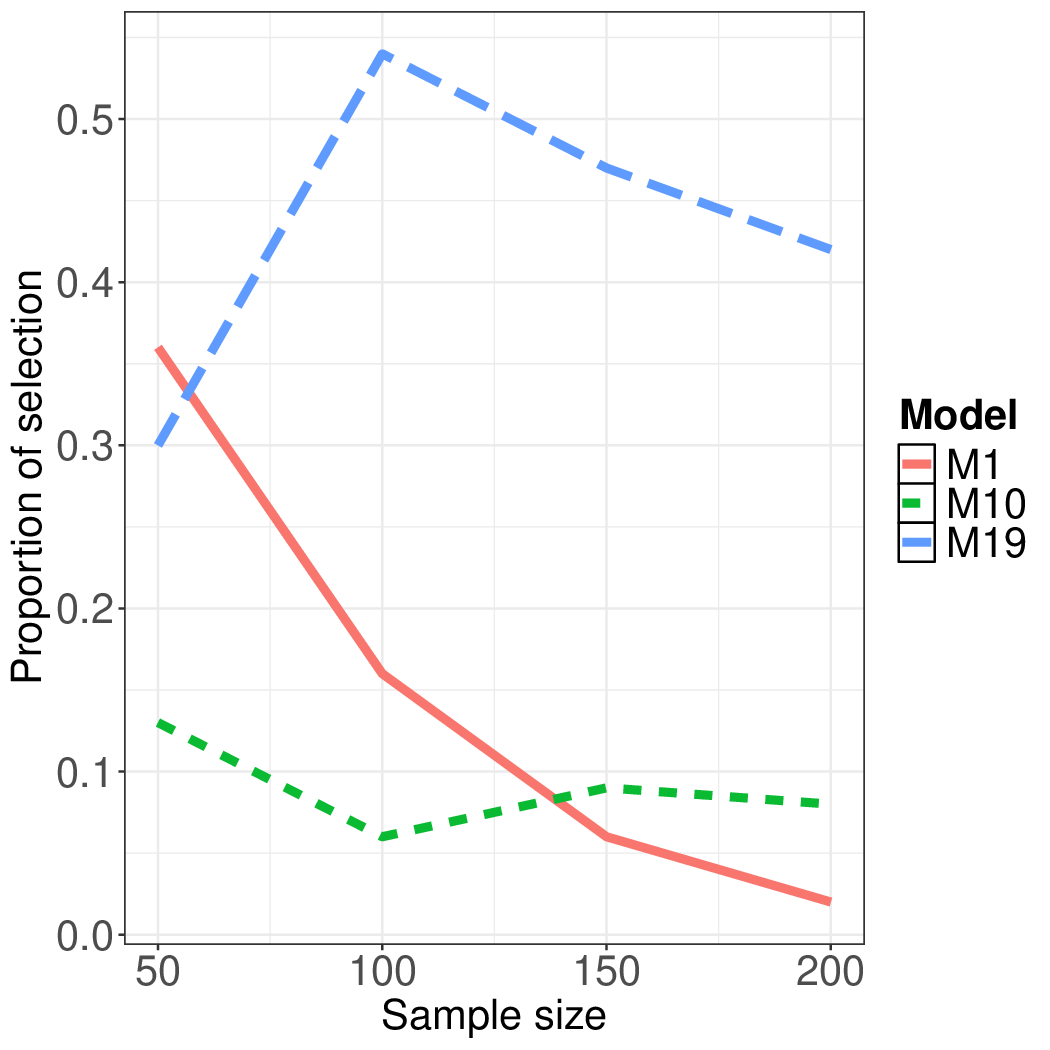}  
		\caption{Selection by AICc}
		\label{fig:sub-AICc-M1}
	\end{subfigure}
	\begin{subfigure}{.5\textwidth}
		\centering
		% include the fourth image
		\includegraphics[width=\linewidth]{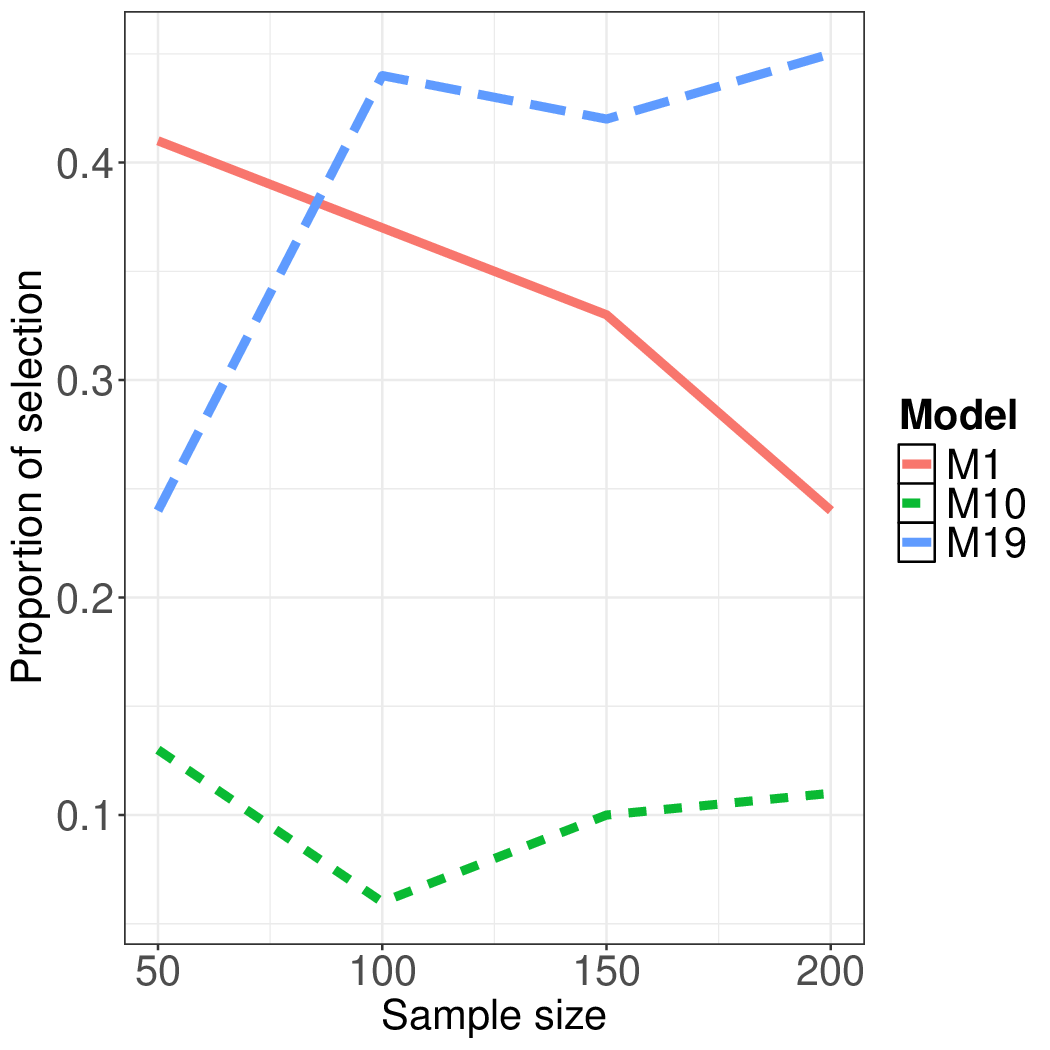}  
		\caption{Selection by BC}
		\label{fig:sub-BC-M1}
	\end{subfigure}
	\caption{ Proportion of times the top three models get selected when the parent model is~M1 by direct optimization}
	\label{fig:fig1}
\end{figure}

\begin{figure}[p]
	\begin{subfigure}{.5\textwidth}
		\centering
		% include the first image
		\includegraphics[width=\linewidth]{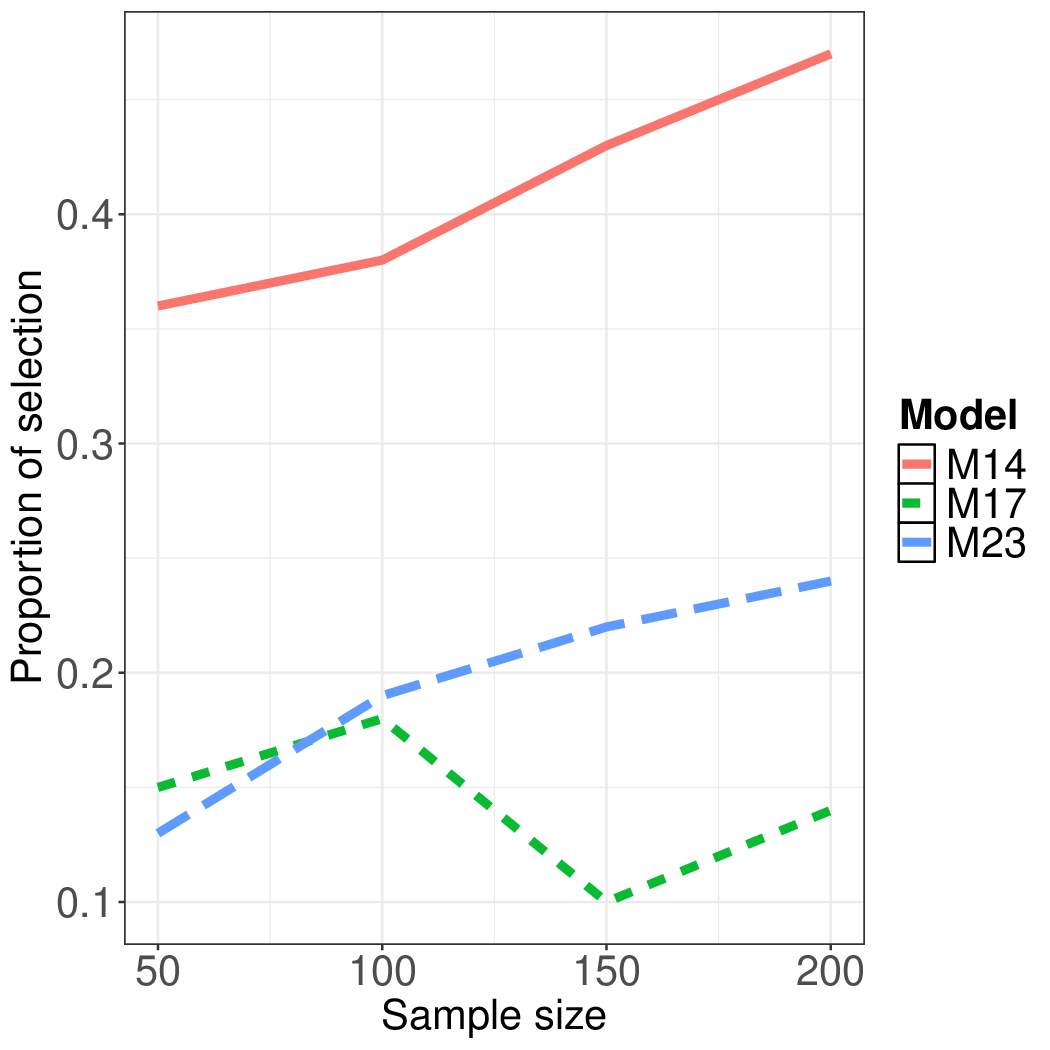}
		\caption{Selection by AIC}
		\label{fig:sub_AIC-M14}
	\end{subfigure}
	\begin{subfigure}{.5\textwidth}
		\centering
		% include the second image
		\includegraphics[width=\linewidth]{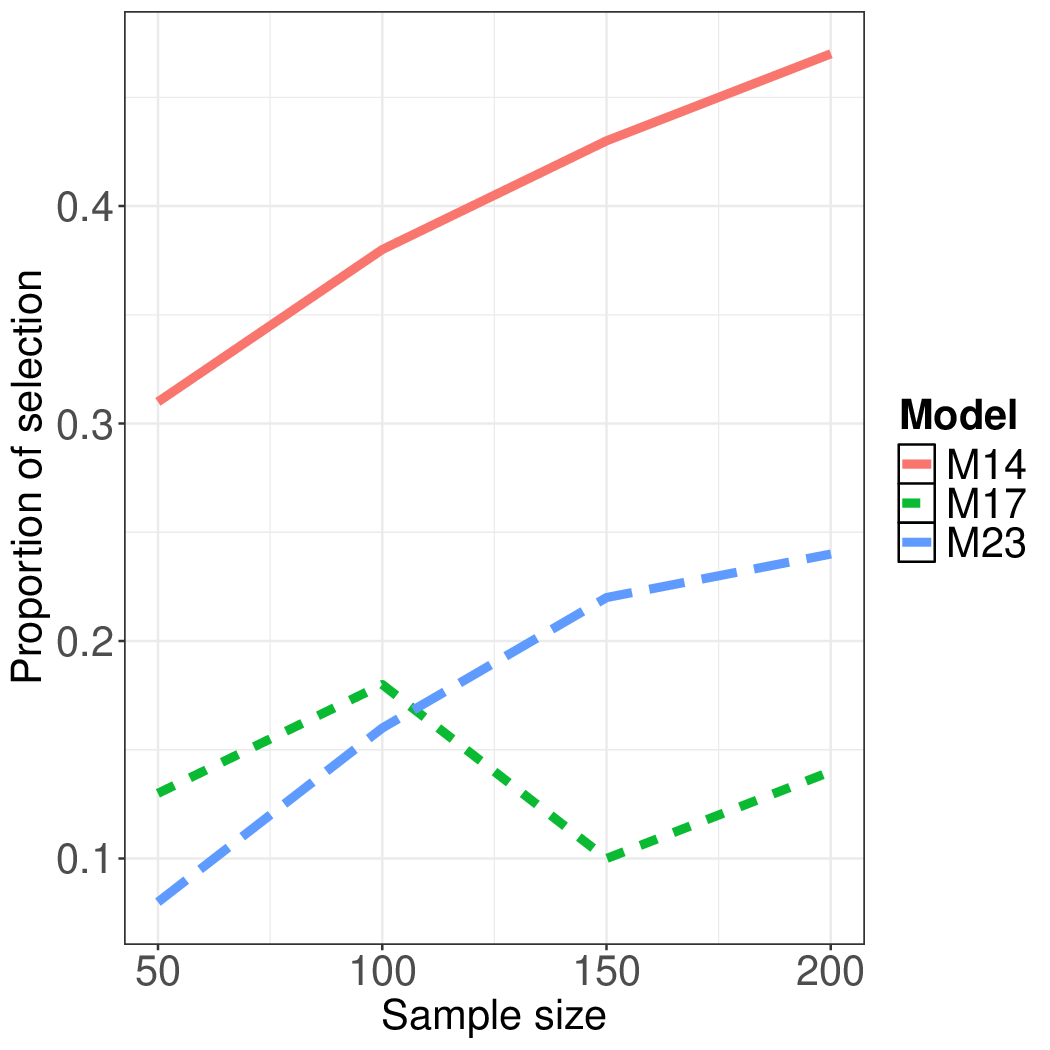}  
		\caption{Selection by BIC}
		\label{fig:sub_BIC-M14}
	\end{subfigure}
	\begin{subfigure}{.5\textwidth}
		\centering
		% include the third image
		\includegraphics[width=\linewidth]{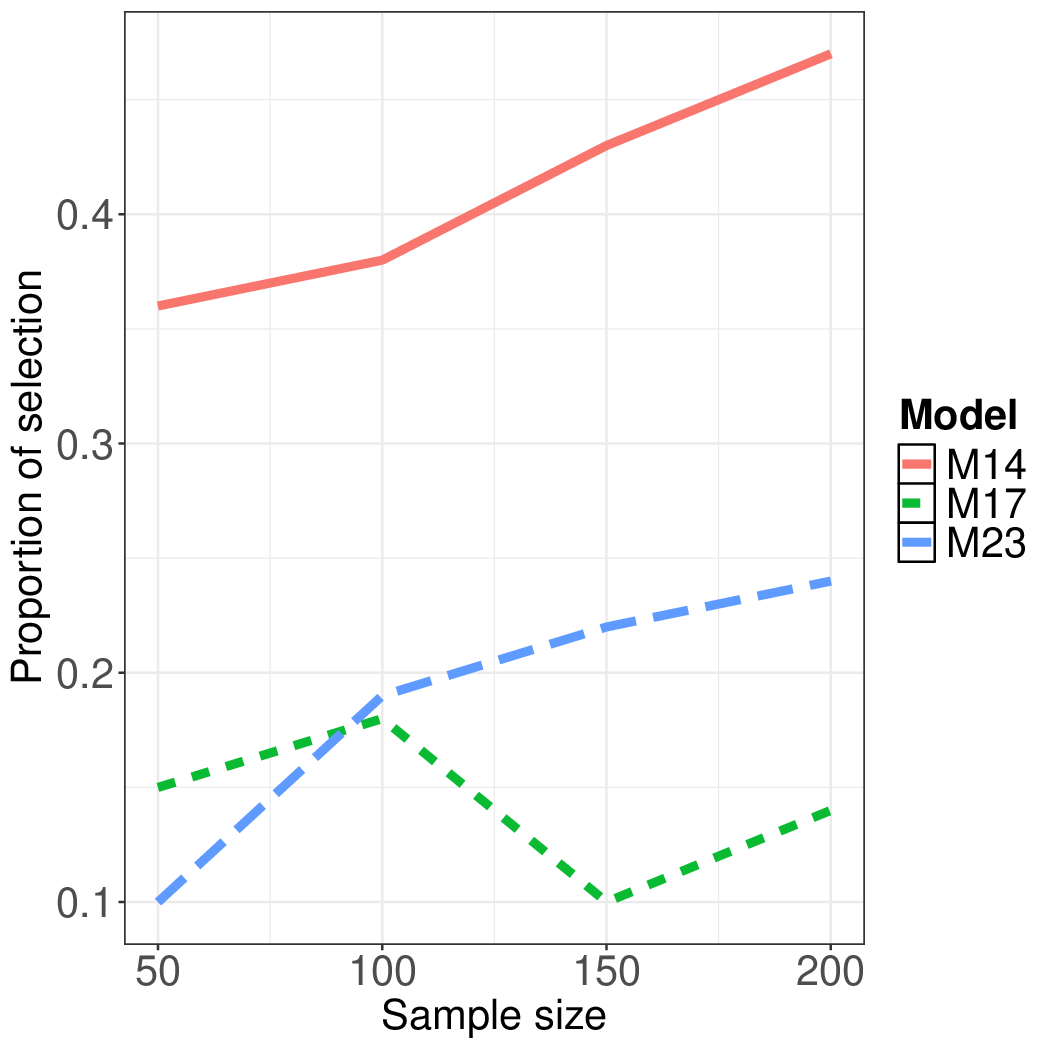}  
		\caption{Selection by AICc}
		\label{fig:sub_AICc-M14}
	\end{subfigure}
	\begin{subfigure}{.5\textwidth}
		\centering
		% include the fourth image
		\includegraphics[width=\linewidth]{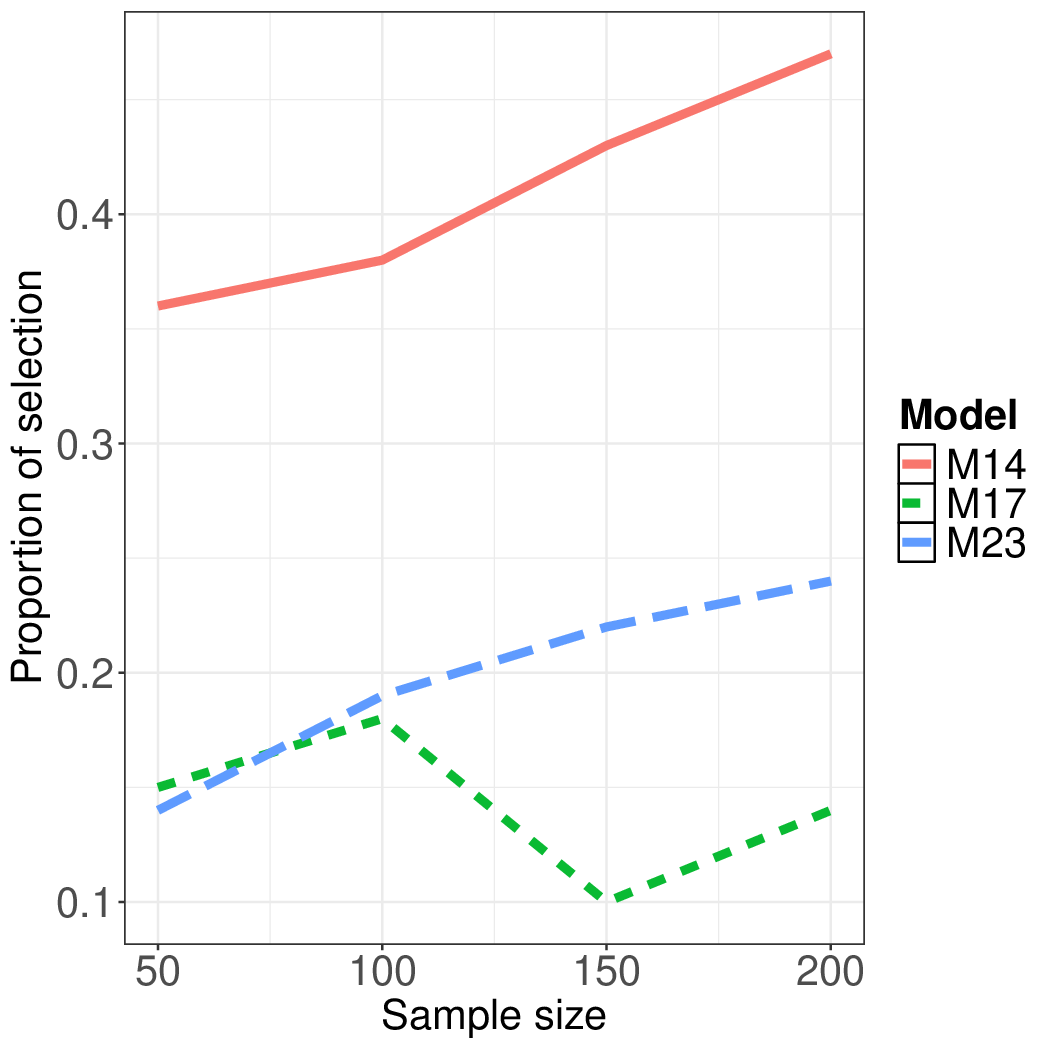}  
		\caption{Selection by BC}
		\label{fig:sub_BC-M14}
	\end{subfigure}
	\caption{ Proportion of times the top three models get selected when the parent model is~M14 by direct optimization}
	\label{fig:fig2}
\end{figure}

As the model selection study is quite elaborate in nature, in this paper we demonstrate a part of the study for illustrative purposes. We generate two-component load-sharing samples from two different parent models, $M1$, and $M14$ for sizes $n=$50, 100, 150, 200. Then, to each generated dataset, we fit all the 27 candidate models $M1$ to $M27$. Figures \ref{fig:fig1} and \ref{fig:fig2} present the proportions of the top three candidate models selected by different model selection criteria.

From Figure \ref{fig:fig1}, it is observed that $M19$ is selected more frequently than the parent model $M1$. This is a plausible case, since $M19$ represents a more general model, with the exponential distribution for the baseline of component lifetimes and the gamma distribution for the frailty, compared to $M1$ which has the exponential distribution for both the component's baselines and the frailty.

From Figure \ref{fig:fig2}, observe that the parent model $M14$ is selected the highest number of times, as one would expect. Also, as the sample size increases, the probability for correct selection steeply increases. These observations clearly establish that the standard model selection criteria can quite accurately identify the parent baseline-frailty combinations for the GFB-GG model structure. 

\subsection{Discussions}
The results of this elaborate simulation study clearly indicate that for two-component load-sharing data, 
\begin{enumerate}
    \item The proposed fitting method can satisfactorily estimate the model parameters of the baseline-frailty distribution combinations within the GFB-GG model structure; 
    \item The standard model selection criteria can accurately identify, even for moderately small sample sizes, the parent distribution with any baseline-frailty combination within the GFB-GG model structure. 
\end{enumerate}

These two findings indicate the principal advantage of using the proposed model: When it is fitted to a given two-component load-sharing data, one can choose the best model involving a specific combination of the baseline and frailty distributions within the GFB-GG model structure. Also, since the GFB and GG families are very flexible, the proposed modeling structure actually provides a broad variety of models to choose from, both for the baseline as well as the frailty distribution. Thus, this structure very effectively captures the dependence between the components of the load-sharing system, which in turn results in suitable inferences for various characteristics related to the system lifetime.         

%%%%%%%%%%%%%%%%%%%%%%%%%%%%%%%%%%%%%%%%%%%%%%%%%%%%
% \begin{figure}[p]
% 	\begin{subfigure}{.5\textwidth}
% 		\centering
% 		% include the first image
% 		\includegraphics[width=\linewidth]{./plots/graph_AIC_M1_13march.eps}
%             \caption{Selection by AIC}
% 		\label{fig:sub-AIC-M1}
% 	\end{subfigure}
% 	\begin{subfigure}{.5\textwidth}
% 		\centering
% 		% include the second image
% 		\includegraphics[width=\linewidth]{./plots/graph_BIC_M1_13march.eps}  
% 		\caption{Selection by BIC}
% 		\label{fig:sub-BIC-M1}
% 	\end{subfigure}
% 	\begin{subfigure}{.5\textwidth}
% 		\centering
% 		% include the third image
% 		\includegraphics[width=\linewidth]{./plots/graph_AICc_M1_13march.eps}  
% 		\caption{Selection by AICc}
% 		\label{fig:sub-AICc-M1}
% 	\end{subfigure}
% 	\begin{subfigure}{.5\textwidth}
% 		\centering
% 		% include the fourth image
% 		\includegraphics[width=\linewidth]{./plots/graph_BC_M1_13march.eps}  
% 		\caption{Selection by BC}
% 		\label{fig:sub-BC-M1}
% 	\end{subfigure}
% 	\caption{\textcolor{blue}{ Proportion of times the top three models get selected when the parent model is~M1 using EM algorithm}}
% 	\label{fig:fig1}
% \end{figure}
%%%%%%%%%%%%%%%%%%%%%%%%%%%%%%%%%%%%%%%%%%%%%%%%
%\textcolor{orange}{ I am also adding the graphs for M1 using direct optimization. Please choose the appropriate one and comment on the other one.}
%%%%%%%%%%%%%%%%%%%%%%%%%%%%%

%%%%%%%%%%%%%%%%%%%%%%%%%%%%%%%%%%%%%%%%%%%%%
%%%%%%%%%%%%%%%%%%%%%%%%%%%%%%%%%%%%%%%%%%%%%%%%%%%%%%
% \begin{figure}[p]
% 	\begin{subfigure}{.5\textwidth}
% 		\centering
% 		% include the first image
% 		\includegraphics[width=\linewidth]{./plots/graph_AIC_M27.eps}  
% 		\caption{Selection by AIC}
% 		\label{fig:sub_AIC-M27}
% 	\end{subfigure}
% 	\begin{subfigure}{.5\textwidth}
% 		\centering
% 		% include the second image
% 		\includegraphics[width=\linewidth]{./plots/graph_BIC_M27.eps}  
% 		\caption{Selection by BIC}
% 		\label{fig:sub_BIC-M27}
% 	\end{subfigure}
% 	\begin{subfigure}{.5\textwidth}
% 		\centering
% 		% include the third image
% 		\includegraphics[width=\linewidth]{./plots/graph_AICc_M27.eps}  
% 		\caption{Selection by AICc}
% 		\label{fig:sub_AICc-M27}
% 	\end{subfigure}
% 	\begin{subfigure}{.5\textwidth}
% 		\centering
% 		% include the fourth image
% 		\includegraphics[width=\linewidth]{./plots/graph_BC_M27.eps}  
% 		\caption{Selection by BC}
% 		\label{fig:sub_BC-M27}
% 	\end{subfigure}
% 	\caption{\textcolor{blue}{Proportion of times the top three models get selected when the parent model is~M27}}
% 	\label{fig:fig3}
% \end{figure}

%%%%%%%%%%%%%%%%%%%%%%%%%%%%%%%%%%%%%%%%%%%%%%%%%%%%%%%%%%
\section{Numerical examples} \label{sec:dataana}
\subsection{A simulation case} 
A simulation case demonstrates the effectiveness of using the proposed GFB-GG model structure for two-component load-sharing systems. The simulated two-component load-sharing dataset is provided in Table~\ref{Sim_data_GG_B6} for interested readers. 

The dataset is of size 200, and is generated using general choices for the baseline and frailty distributions within the GFB-GG model structure. The baseline distributions of the component lifetimes, before the first failure, are taken to be Weibull within the GFB family. After the first failure occurs, the lifetime of the remaining component is generated from Gamma distribution that also belongs to the GFB family. The shared frailty that connects the two components is generated from the GG distribution with general parameter choices, i.e., these parameter values for the GG distribution do not lead to any special member within the GG family of distributions. Specifically, the true parameter values are taken to be $\theta_1=0.25, ~\theta_2=0.50,~ \theta_1^*=0.75, ~\theta_2^*=1,~ \theta_B=2, ~\theta_B^*=1.5,~ k=1.5, ~\beta=1.5,~ \sigma=1.5$. 

 To this simulated data, a total 45 models with various baseline-frailty distribution combinations are fitted. In addition to the 27 model combinations presented in Table \ref{tab:modelname}, another 18 baseline-frailty model combinations (9 each for lognormal and generalized gamma frailty combined with different baselines from the GFB model) are fitted to the simulated data by using the proposed fitting method.Table~\ref{tab:simulated_data_analysis} presents the top three models among them, according to their AIC values. The table also provides the standard errors, and lower and upper limits of 95\% bootstrap confidence intervals.

We observe that the model with Weibull-Gamma baselines within the GFB distribution, and the GG frailty is selected as the best model for this data, as this model has the lowest AIC value among the 45 candidate models. Recall that this combination was the true parent model too. Note that, instead of fitting the general GFB-GG model structure with 45 candidate models to this two-component load-sharing data, if any special combination of baseline-frailty (such as, the model of Asha et al.~\cite{AKK2016}) was fitted, it would have resulted in a sub-optimal model fit. Only when the GFB-GG model structure is fitted to this data, and a model selection is performed within the families, one can identify the best model for this data. This clearly demonstrates the utility of the proposed GFB-GG model structure for two-component load-sharing data in obtaining the optimal model fit.

\begin{table}[ht] \scriptsize
	\centering
	\caption{Simulated two-component load-sharing data: Baseline component lifetimes as Weibull($\theta_B=2$) before the first failure, and Gamma($\theta_B^*=1.5$) after the first failure, with PFR parameters ($\theta_1=0.25,~ \theta_2=0.5, ~\theta_1^*=0.75,~ \theta_2^*=1$); frailty from GG distribution ($k=1.5,~\beta=1.5$).}
% \begin{table}[ht]
% \centering
\begin{tabular}{|c|c|c|c|c|c|c|c|c|c|c|c|c|c|c|c|}
  \hline
$n$ & $Y_1$& $Y_2$ & $n$ & $Y_1$& $Y_2$ & $n$ & $Y_1$& $Y_2$ & $n$ & $Y_1$& $Y_2$ & $n$ & $Y_1$& $Y_2$\\ 
  \hline
    1 & 0.55 & 0.44 & 2 & 5.77 & 1.30 & 3 & 1.86 & 0.77 & 4 & 3.66 & 0.31 &5 & 4.51 & 1.64 \\
    6 & 1.15 & 4.30 & 7 & 0.79 & 2.96 & 8 & 1.89 & 1.87 & 9 & 0.83 & 1.37 & 10 & 1.36 & 1.27 \\
    11 & 3.90 & 1.35 & 12 & 2.52 & 1.37 & 13 & 1.83 & 4.37 & 14 & 1.95 & 0.88 & 15 & 1.10 & 1.89\\
    16 & 1.07 & 1.19 & 17 & 0.67 & 0.22 & 18 & 2.69 & 3.04 & 19 & 1.74 & 1.21 & 20 & 0.61 & 3.57 \\ 
    21 & 1.22 & 1.14 & 22 & 1.77 & 1.28 &  23 & 1.00 & 1.19 & 24 & 3.86 & 1.02 & 25 & 1.14 & 0.86 \\
    26 & 1.30 & 3.42 & 27 & 1.70 & 1.04 & 28 & 3.05 & 0.57 & 29 & 4.70 & 1.04 &  30 & 3.79 & 0.62 \\
    31 & 1.06 & 0.57 & 32 & 3.30 & 4.94 & 33 & 0.44 & 1.18 &  34 & 0.70 & 2.70 & 35 & 4.96 & 2.62 \\
    36 & 1.73 & 1.68 & 37 & 11.32 & 0.94 & 38 & 2.68 & 0.53 & 39 & 1.67 & 1.06 & 40 & 1.18 & 5.93 \\ 
   41 & 0.89 & 1.43 & 42 & 1.67 & 0.65 & 43 & 5.70 & 2.70 & 44 & 10.17 & 2.17 & 45 & 3.61 & 1.04 \\ 
   46 & 1.29 & 7.00 & 47 & 0.22 & 2.08 & 48 & 2.11 & 6.11 & 49 & 10.95 & 0.95 &  50 & 1.11 & 1.77 \\ 
   51 & 1.11 & 0.95 & 52 & 0.58 & 1.30 & 53 & 1.77 & 2.88 & 54 & 2.13 & 1.47 & 55 & 5.31 & 1.60 \\ 
   56 & 2.74 & 1.47 & 57 & 1.77 & 1.48 &  58 & 2.57 & 1.29 & 59 & 1.55 & 1.28 & 60 & 0.37 & 1.34 \\ 
   61 & 9.32 & 1.83 & 62 & 0.19 & 1.25 &  63 & 9.55 & 4.45 &  64 & 1.09 & 0.63 & 65 & 1.86 & 1.45 \\ 
   66 & 1.16 & 0.61 &  67 & 1.43 & 0.95 &  68 & 4.38 & 0.24 &  69 & 0.56 & 1.80 &  70 & 0.79 & 0.37 \\ 
   71 & 8.57 & 4.46 & 72 & 1.39 & 5.69 & 73 & 2.23 & 0.52 & 74 & 2.88 & 0.83 & 75 & 0.98 & 1.56 \\ 
   76 & 1.15 & 3.02 & 77 & 36.40 & 4.03 & 78 & 2.71 & 3.36 & 79 & 0.82 & 1.24 &  80 & 12.17 & 0.66 \\ 
   81 & 0.72 & 0.79 & 82 & 0.93 & 1.01 & 83 & 2.51 & 0.54 & 84 & 0.66 & 0.69 & 85 & 2.19 & 0.73 \\ 
   86 & 1.73 & 8.69 & 87 & 3.25 & 0.19 & 88 & 2.62 & 1.13 & 89 & 1.79 & 1.42 & 90 & 1.29 & 3.70 \\ 
   91 & 0.96 & 1.06 & 92 & 0.70 & 0.31 & 93 & 1.13 & 1.21 & 94 & 0.95 & 3.19 & 95 & 8.02 & 0.54 \\ 
   96 & 1.11 & 2.12 & 97 & 2.23 & 2.23 & 98 & 1.50 & 0.56 & 99 & 1.39 & 0.22 & 100 & 2.66 & 1.30 \\ 
  101 & 0.52 & 1.45 & 102 & 4.96 & 1.58 & 103 & 0.94 & 0.72 & 104 & 0.34 & 1.84 & 105 & 7.91 & 1.77 \\ 106 & 2.61 & 0.33 & 107 & 1.82 & 0.90 & 108 & 3.35 & 1.31 & 109 & 5.00 & 0.28 & 110 & 1.85 & 8.25 \\ 
  111 & 1.73 & 4.73 & 112 & 2.39 & 1.12 & 113 & 3.18 & 0.88 & 114 & 0.93 & 1.57 & 115 & 1.45 & 3.94 \\ 
  116 & 0.65 & 0.24 & 117 & 0.99 & 1.55 & 118 & 4.16 & 1.29 & 119 & 1.15 & 0.87 & 120 & 0.50 & 0.53 \\ 
  121 & 0.45 & 0.76 & 122 & 0.83 & 2.19 & 123 & 1.63 & 0.62 & 124 & 0.54 & 0.22 & 125 & 1.80 & 0.65 \\ 
  126 & 6.89 & 0.57 & 127 & 5.72 & 1.50 & 128 & 4.04 & 1.38 & 129 & 0.94 & 4.43 & 130 & 3.08 & 0.88 \\ 
  131 & 3.12 & 2.63 & 132 & 0.95 & 0.85 & 133 & 1.44 & 1.69 & 134 & 0.52 & 5.25 & 135 & 0.55 & 1.04 \\ 
  136 & 0.90 & 2.12 & 137 & 2.47 & 1.53 & 138 & 1.24 & 2.07 & 139 & 2.43 & 1.49 & 140 & 1.44 & 0.53 \\
  141 & 0.63 & 1.06 & 142 & 0.92 & 2.88 & 143 & 1.35 & 0.87 & 144 & 1.10 & 0.68 & 145 & 0.54 & 3.58 \\ 
  146 & 2.39 & 0.21 & 147 & 1.76 & 0.08 & 148 & 1.94 & 1.24 & 149 & 3.51 & 0.99 & 150 & 2.13 & 0.75 \\ 
  151 & 2.19 & 1.40 & 152 & 0.91 & 1.64 & 153 & 1.62 & 1.78 & 154 & 0.96 & 0.86 & 155 & 5.73 & 0.98 \\ 
  156 & 3.82 & 1.72 & 157 & 4.51 & 1.17 & 158 & 2.12 & 3.65 & 159 & 1.43 & 2.61 & 160 & 1.76 & 1.44 \\ 
  161 & 3.48 & 1.69 & 162 & 0.91 & 0.73 & 163 & 0.64 & 6.09 & 164 & 0.04 & 0.60 & 165 & 0.77 & 1.95 \\ 
  166 & 8.23 & 2.24 & 167 & 1.31 & 2.63 & 168 & 7.52 & 1.04 & 169 & 4.62 & 0.65 & 170 & 0.37 & 1.09 \\ 
  171 & 1.19 & 0.77 & 172 & 0.10 & 1.01 & 173 & 2.30 & 2.39 & 174 & 1.25 & 0.59 & 175 & 1.08 & 1.17 \\ 
  176 & 6.64 & 1.47 & 177 & 3.35 & 0.52 & 178 & 1.09 & 1.04 & 179 & 1.53 & 2.04 & 180 & 5.59 & 0.98 \\ 
  181 & 0.17 & 0.65 & 182 & 1.66 & 0.62 & 183 & 2.52 & 0.95 & 184 & 1.12 & 1.27 & 185 & 3.76 & 0.75 \\ 
  186 & 1.39 & 0.68 & 187 & 1.75 & 0.47 & 188 & 1.00 & 1.81 & 189 & 3.18 & 3.06 & 190 & 1.94 & 0.89 \\ 
  191 & 2.87 & 1.08 & 192 & 2.10 & 0.98 & 193 & 1.20 & 0.88 & 194 & 1.83 & 1.36 & 195 & 1.74 & 1.43 \\ 
  196 & 1.06 & 0.66 & 197 & 5.53 & 2.09 & 198 & 1.83 & 0.72 & 199 & 3.84 & 0.94 & 200 & 1.10 & 0.48\\ 
   \hline
\end{tabular}\label{Sim_data_GG_B6}
\end{table}

\begin{table}[p] \scriptsize
	\centering
	\caption{Estimates of the parameters of top three models based on simulated data given in Table~\ref{Sim_data_GG_B6}.}
	\begin{tabular}{c|c|cccccccccc}
		\toprule
		Model   &  Parameters & $\theta_1$  & $\theta_2$  & $\theta_1^*$ & $\theta_2^*$& $\theta_B $ & $\theta_B^*$ & $k$ & $\beta$ & $\sigma$  & AIC\\
		%\cmidrule(lr){2-10}	
		% Number & True Value & 0.3 & 0.4 & 0.5 & 1.0 & 2.0 & 1.5 & 1.5\\		
		\midrule
		 &MLE & 0.317&	0.539&	0.871&	1.251&2.147 &1.820&		1.415& 1.627 &-& 1278.079 \\
		\cmidrule(lr){2-12}
        $R_B$: Weibull($\theta_B$,1)&Boot SE & 	0.015 &	0.022&	0.084&	0.150&	0.051 &	0.156 &0.342&	0.672 &- &\\
		\cmidrule(lr){2-12}
		$R_B^*$: Gamma($\theta_B^*$,1) & Boot LL & 	0.296&	0.503&	0.735&	0.935&2.064&	1.585&	0.990&	0.412&-	 &\\
		\cmidrule(lr){2-12}
		Frailty: GG($\frac{\Gamma(\beta)}{\Gamma((\beta+\frac{1}{\beta})},~k,~ \beta$) & Boot UL &  	0.354&	0.588&	1.064&	1.524& 2.264&	2.196&	2.332&	3.047&-	  &\\

            \midrule
		&MLE & 	0.350&	0.596&	1.132&	1.674&2.205&	2.127&	-& - & 0.746&1278.326 \\
		\cmidrule(lr){2-12}
		%LN-Weibull-Gamma 
        $R_B$: Weibull($\theta_B$,1)&Boot SE & 	0.017&	0.031&	0.111&	0.186& 0.056&	0.130&-&-&	0.047	 &\\
		\cmidrule(lr){2-12}
		$R_B^*$: Gamma($\theta_B^*$,1)& Boot LL & 	0.314&	0.531&	0.921&	1.291&2.102&	1.845&-&-&	0.649	 &\\
		\cmidrule(lr){2-12}
		Frailty: Log-normal($-\frac{\sigma^2}{2}$, $\sigma^2$)& Boot UL & 	0.382& 0.653&	1.357&	2.020&2.321&	2.355&-&-&	0.834	 &\\

            \midrule
		&MLE & 	 0.314&	 0.535&	  0.399&	0.572 & 2.153 & 1.258 & 1.862 &1.087& -& 1278.590\\
		\cmidrule(lr){2-12}
		%GG-Weibull-Weibull
       $R_B$: Weibull($\theta_B$,1) &Boot SE & 0.017&	0.020&	0.028&	0.036&0.051&	0.052&	0.336&	0.296&	 -&\\
		\cmidrule(lr){2-12}
		$R_B^*$: Weibull($\theta_B^*$,1)& Boot LL & 	0.292&	0.492&	0.342&	0.502&2.095&	1.158&	1.378&	0.399&-	 &\\
		\cmidrule(lr){2-12}
		Frailty: GG($\frac{\Gamma(\beta)}{\Gamma((\beta+\frac{1}{\beta})},~k,~ \beta$)& Boot UL & 	0.357&	0.569&	0.450&	0.641&2.296&	1.363&	2.697&	1.561&-	 	 &\\

		\bottomrule

	\end{tabular}
		\label{tab:simulated_data_analysis}
\end{table}

\subsection{Real data: The nuclear reactor data}
The dataset on nuclear reactors, obtained from 4 nuclear sites in South Korea, was analyzed by Park and Kim~\cite{ParkKim2014}. The data consists of 30 pairs of failure times ($Y_1$) and warranty servicing times($Y_2$) of 20 operational nuclear power plants across all the nuclear sites. The dataset is presented here in Table~\ref{tab:realdata}; See Park and Kim~\cite{ParkKim2014} for more details on the dataset. Park and Kim's analysis was aimed at cost analysis on warranty policies for the nuclear reactors, for which they used Freund's bivariate exponential model~\cite{Freund1961}. Through nonparametric analysis, they concluded that failure times and warranty servicing times were dependent variables, and their warranty model included both these variables as significant factors. Very recently, Franco et al.~\cite{FVK2020} have also analyzed the dataset. In Franco et al.'s analysis, the stochastic dependence between failure times and warranty servicing times has been established. It has also been observed that the warranty servicing times are strongly influenced by the overload due to the failure times of the nuclear reactors~\cite{FVK2020}.

In this section, we present an analysis of the dataset on nuclear reactor data, by using the proposed GFB-GG model. All the 45 models with various baseline-frailty distribution combinations are fitted to the data, and model selection within the family has been performed in a similar fashion as explained in Section \ref{subsec:modelselection}. The failure times of the nuclear plants are scaled before analysis, by dividing them by 365. This will not affect the inference in any way. The top three models for the nuclear reactor data, in order of lowest AIC values, are presented in Table~\ref{tab:realdata_analysis}, along with the point estimates, standard errors of estimates, and bootstrap confidence intervals of their parameters. 

The best model, according to AIC, is the one with exponential-gamma baseline, and Weibull frailty, with an AIC value of -175.469 which is the lowest among the 45 models considered. It may also be mentioned here that for the best model for this data reported in Franco et al.~\cite{FVK2020}'s work has an AIC value -172.342. Recall that Franco et al. also considered the GFB distribution for the baseline component lifetimes, but they did not consider a frailty that induced dependence due to unobserved random factors. Comparing the AIC values, we can conclude that the best model for the nuclear reactor data within the GFB-GG model structure provides a better fit compared to the best model of Franco et al.~\cite{FVK2020}. This once again shows the usefulness of the proposed model. 

\begin{table}[ht]\scriptsize
         \centering
	\caption{The nuclear reactor dataset reported in Park and Kim~\cite{ParkKim2014}}
	%\vspace{0.2cm}
	\begin{tabular}{|c|c|c|c|c|c|c|c|c|c|c|c|c|c|c|c|}
  \hline
$n$ & $Y_1$& $Y_2$ & $n$ & $Y_1$& $Y_2$ & $n$ & $Y_1$& $Y_2$ & $n$ & $Y_1$& $Y_2$ & $n$ & $Y_1$& $Y_2$\\ 
  \hline
 1 &	353.04	& 4.37 & 2	& 334.72	& 1.91 & 3	& 80.04	& 2.04 & 4	& 6.49	& 1.72 &  5	& 1.34 & 0.29\\
 6 & 467.19 & 1.93 & 7  &	0.35	& 1.82 & 8 &	398.86	& 1.77 & 9	& 1048.23 &	9.61 & 10 & 829.39 & 3.80\\
11 &	227.20 &	2.86 & 12 &	260.14	& 0.31 & 13 & 	14.00	& 0.85 & 14	& 14.15	& 2.04 & 15	& 38.96	& 2.73\\
16	& 30.27	& 3.63 & 17	& 117.37	& 2.73 & 18	& 126.27	& 2.55 & 19	& 56.45	& 0.72 & 20	& 45.28	& 3.69\\
21	& 267.31	& 0.36 & 22	& 615.64 &	10.63 & 23 &	115.37 &	11.24 & 24	& 359.76	& 9.70 & 25 &	412.30 &	3.31 \\
26	& 276.69 &	4.96 & 27	& 601.04	& 2.99 & 28	& 1021.01	& 2.36 & 29	& 192.17	& 1.63 & 30	& 0.36	& 0.26\\
\hline
\end{tabular}	\label{tab:realdata}
\end{table}

\begin{table}[p] \scriptsize
	\centering
	\caption{Estimates of the parameters of top three models based on the nuclear reactor data given in Table~\ref{tab:realdata}}
	\begin{tabular}{c|c|ccccccccc}
		\toprule
		Model  &  Parameters & $\theta_1$  & $\theta_2$  & $\theta_1^*$ & $\theta_2^*$& $\theta_B $ & $\theta_B^*$ & $k$ & $\beta$ & AIC\\
		%\cmidrule(lr){2-10}	
		% Number & True Value & 0.3 & 0.4 & 0.5 & 1.0 & 2.0 & 1.5 & 1.5\\		
		\midrule
		&MLE & 	3.751 &	108.749 &	0.663 &	12.674 &-&	0.379 &	110.001 &-&-175.469\\
		\cmidrule(lr){2-11}
		$R_B$: Exp(1)&Boot SE &	0.590 &	4.978 &	0.061 &	0.109 &	-& 0.057&0.003 &-&\\
		\cmidrule(lr){2-11}
		$R_B^*$: Gamma($\theta_B^*$,1)& Boot LL  &	2.573 &	99.488&	0.516 &	12.482 &-&	0.242 &109.990	&-  &\\
		\cmidrule(lr){2-11}
		Frailty: Weibull($k$, $\frac{1}{\Gamma(1+\frac{1}{k})}$)& Boot UL &	4.887&	119.002&	0.754 &	12.910 &-& 0.465 &	110.002	 &- &\\

            \midrule
		&MLE & 3.765&	109.212& 0.666&	 12.700 &-&	0.382	& 	-&273.502&	-175.464\\
		\cmidrule(lr){2-11}
		$R_B$: Exp(1)&Boot SE & 	0.548&	5.495&	0.056&	0.135&-&0.063&-&	0.017	 &\\
		\cmidrule(lr){2-11}
		$R_B^*$: Gamma($\theta_B^*$,1)& Boot LL & 	2.635&	98.154&	0.525& 12.462&-& 0.264&	-&	273.452	 &\\
		\cmidrule(lr){2-11}
		Frailty: Gamma($\beta$, $\frac{1}{\beta}$)& Boot UL & 	4.785&	119.696&	0.743&12.989&-&	0.510&-&273.518	  &\\

            \midrule
		&MLE & 	 8.000&	226.000&	0.761&	15.000& 1.148&	0.440&	-&12.000 &	-174.270	 \\
		\cmidrule(lr){2-11}
		$R_B$: Weibull($\theta_B$,1)&Boot SE & 	1.553&	1.587&	0.079&	0.054&0.019&	0.065&	-&4.838	 &\\
		\cmidrule(lr){2-11}
		$R_B^*$: Gamma($\theta_B^*$,1)& Boot LL & 		4.591&	222.549&	0.619&	14.900&	1.120&	0.326&-&5.866	 &\\
		\cmidrule(lr){2-11}
		Frailty: Gamma($\beta$, $\frac{1}{\beta}$)& Boot UL & 	10.679&	228.769&	0.928&	15.111&1.195&	0.581&	-&24.829	 &\\

  %           \midrule
		% &MLE & 	1.12&	0.40&	6.70&	195.00&	0.68&	13.80&	64.00&	-174.081	\\
		% \cmidrule(lr){2-10}
		% M27&SE & 	0.08&	0.22&	8.14&	66.96&	0.27&	21.48&	25.69	&\\
		% \cmidrule(lr){2-10}
		% & Boot LL & 	0.98&	0&	0&	62.93&	0.07&	0&	12.36&\\
		% \cmidrule(lr){2-10}
		% & Boot UL & 1.29&	0.79&	22.84&	325.41&	1.11&	46.63&	113.08&\\

  %          \midrule
		% &MLE & 	1.09&	0.39&	5.80&	165.89&	0.68&	13.12&	10.98&	-173.997	\\
		% \cmidrule(lr){2-10}
		% M18&SE & 	0.08&	0.22&	8.14&	66.96&	0.27&	21.48&	25.69	&\\
		% \cmidrule(lr){2-10}
		% & Boot LL & 	0.98&	0&	0&	62.93&	0.07&	0&	12.36&\\
		% \cmidrule(lr){2-10}
		% & Boot UL & 1.29&	0.79&	22.84&	325.41&	1.11&	46.63&	113.08&\\        

		\bottomrule

	\end{tabular}
		\label{tab:realdata_analysis}
\end{table}	
\section{Conclusions} \label{sec:con}

In this paper, we have presented a very general class of models for two-component load-sharing systems. The proposed model with the GFB and GG family of distributions for the baseline component lifetimes and frailty distributions, respectively, can appropriately model the change in the lifetime distribution of the surviving component after the first failure occurs in the system, and can also accurately capture the dependence between the component lifetimes. A model fitting method based on an EM-type algorithm is described in detail, and extensive simulations are carried out. The fitting method is observed to perform satisfactorily. It is shown that the GFB-GG model structure can provide a better fit compared to existing models as one can choose from a wider array of candidate models in this case. This demonstrates the practical utility of the proposed model.    

\section*{\sc Funding information}
\begin{itemize}
\item The research of Debanjan Mitra is supported by the Mathematical Research Impact
Centric Support (File no.~MTR/2021/000533) from the Science and
Engineering Research Board, Department of Science and Technology, Government of
India.
\end{itemize}

\end{document}